\documentclass[10pt,twocolumn,twoside]{IEEEtran}
\usepackage{cite}
\usepackage{amsmath,amssymb,amsfonts}
\usepackage{graphicx}
\usepackage{textcomp}
\usepackage{algorithm}
\usepackage{subcaption}
\usepackage{multirow}
\usepackage{enumerate}
\usepackage{color}

\usepackage{amsthm}
\usepackage{setspace}

\usepackage{algpseudocode}
\usepackage{array}
\floatname{algorithm}{Algorithm}

\DeclareMathOperator*{\argmin}{arg\,min}

\DeclareMathOperator*{\maximize}{maximize}
\newcommand{\paren}[1]{\mathopen{}\left( {#1}_{{}_{}}\,\negthickspace\right)\mathclose{}}

\graphicspath{ {./Figs/} }

\newtheoremstyle{definition}
  {\topsep}   
  {\topsep}   
  {\normalfont}  
  {}           
  {\bfseries}  
  {.}          
  {.5em}       
  {}           
\theoremstyle{definition}
\newtheorem{definition}{Definition}
\newtheorem{remark}{Remark}

\newtheorem{lemma}{Lemma}
\newtheorem{assumption}{Assumption}

\newtheorem{theorem}{Theorem}
\theoremstyle{plain}

 \def\BibTeX{{\rm B\kern-.05em{\sc i\kern-.025em b}\kern-.08em
     T\kern-.1667em\lower.7ex\hbox{E}\kern-.125emX}}

\begin{document}

\title{Model-Free Dynamic Consensus in Multi-Agent Systems: A Q-Function Perspective}
\author{Maryam Babazadeh, \IEEEmembership{Senior Member IEEE}, and Naim Bajcinca
\thanks{
Maryam Babazadeh (Corresponding author) is with the Department of Mechanical and Process Engineering, University of Kaiserslautern-Landau (RPTU), Kaiserslautern, Germany (email: m.babazadeh@rptu.de).\\ }}

\maketitle

 \begin{abstract}
This paper presents a new method for dynamic consensus in linear discrete-time homogeneous multi-agent systems (MAS). Achieving state consensus in such systems involves constraints linked to the graph’s spectral properties, complicating the design of coupling gains, especially in large-scale networks. The proposed approach reformulates the dynamic consensus problem with a prescribed convergence rate by introducing a state–action value function within a synthetic linear quadratic regulation (LQR) framework, thereby expressing the problem as a semidefinite program (SDP). The resulting SDP supports the joint design of local feedback and coupling gains in both model-based and model-free settings. To handle non-convex feasibility conditions, a convex–concave decomposition strategy is developed, guaranteeing convergence to a stationary point. In the fully model-free case, the method eliminates the need for system identification or explicit knowledge of agent dynamics, relying solely on input–state data to construct an equivalent data-driven SDP. Finally, a new algorithm balancing feasibility, convergence rate, and energy efficiency enhances design flexibility. Numerical results demonstrate the effectiveness of the proposed method in diverse scenarios.

 \end{abstract}

\section{Introduction}\label{sec:introduction}

The problem of dynamic consensus in multi-agent systems has received significant attention due to its relevance in various domains, including cooperative robotics, sensor networks, formation control,  social networks, synchronization, and collective decision making in decentralized systems \cite{qin2016recent,xia2015structural}. At its core, dynamic consensus entails the ability of agents to reach a common agreement through local interactions by a communication network \cite{liang2019prescribed,chen2019control}.

Consensus problems are fundamentally linked to the communication graph structure, as its connectivity governs the feasibility and convergence of consensus algorithms. For undirected graphs, connectivity suffices, whereas directed graphs require a spanning tree \cite{zhang2011lyapunov}. In continuous-time systems, consensus is achievable if agent dynamics are controllable and the graph contains a spanning tree \cite{li2009consensus}. Strict Lyapunov functions for synchronization errors of homogeneous continuous-time MAS are proposed in \cite{dutta2022strict}, along with a time-varying, strictly increasing coupling gain that removes the need to know the Laplacian matrix. In discrete-time systems, consensusability is more restrictive due to additional constraints and bounded consensus regions \cite{li2009consensus,zhang2020decoupling}, which arise from the graph’s spectral properties and complicate the design of feedback and coupling gains, particularly for large-scale systems with arbitrary Laplacian spectra.
In \cite{feng2021consensusability} a unified approach for dynamic consensus of homogeneous discrete-time MAS is proposed which identifies the maximal disk-guaranteed gain margin (GGM) of discrete-time LQR. Using the inverse optimal control framework, the authors show that achieving globally optimal consensus is possible if and only if the Laplacian matrix of the system is simple and all its non-zero eigenvalues lie within a specific subset of the consensus region when radially projected.

In many practical scenarios, accurate mathematical models of dynamical systems are unavailable. To overcome this limitation, data-driven control has emerged as a powerful paradigm that designs controllers directly from input–output data, bypassing explicit system identification. Using persistently exciting (PE) input sequences, stabilization and optimality of centralized linear systems can be achieved without knowledge of system matrices \cite{de2019formulas}. This idea has been further developed in several studies \cite{berberich2020trajectory, trentelman2021informativity, van2020noisy, farjadnasab2022model, Sahel}, which advance both the theoretical and practical foundations of data-driven control. Extensions to distributed and multi-agent settings have also been explored. For example, \cite{yang2020cooperative} proposed a distributed adaptive data-driven control scheme for unknown single-input single-output (SISO) agents, assuming minimum-phase dynamics with known relative degrees. Subsequent works \cite{baldi2020distributed, cao2023distributed} relaxed these assumptions to handle non-minimum phase dynamics with unknown relative degrees, though they remain limited to SISO systems. More recently, \cite{jiang2024fully} introduced a fully distributed output consensus approach for uncertain homogeneous MAS, using the Laplacian estimation method of \cite{gusrialdi2017distributed} to design both state- and output-feedback controllers. A related study \cite{zhang2023data} proposed a data-driven consensus design for homogeneous MAS that learns only the feedback gains; however, it assumes a zero control-weighting matrix in the Riccati-like equation and neglects feasibility constraints induced by the graph Laplacian spectrum. In \cite{feng2021q}, a Q-learning algorithm was applied to model-free consensus in homogeneous MAS with unknown agent dynamics, but it focuses on single-input agents and depends on the availability of an initial stabilizing policy, either obtained heuristically or from expert knowledge.

This paper introduces a novel approach for achieving dynamic consensus in linear discrete-time homogeneous multi-agent systems.
While prior studies on dynamic consensus, such as \cite{feng2021consensusability}, have established conservative conditions for achieving consensus, they leave open the challenge of developing a systematic approach to solve the underlying feasibility problem or to improve the feasibility properties of these conditions. Such challenges become particularly pronounced for directed networks with complex-conjugate eigenvalues or for large-scale systems where multiple coupled constraints must be satisfied and the designed gains must remain consistent across all agents to ensure coherent global behavior. In contrast, the proposed approach addresses this gap through a constructive, optimization-based framework that formulates new feasibility conditions, enlarges the feasible region, and directly computes feasible controller and coupling gains, making it applicable to large and directed networks. Building on the idea of parameterizing the LQR problem in dual form via Q-function elements as proposed in \cite{farjadnasab2022model}, we reformulate the eigenvalue-dependent feasibility conditions in terms of Q-function parameters. This novel integration provides a fully model-free dynamic consensus scheme that can be realized from a single persistently exciting data trajectory. The main contributions of this work are as follows: (i) To the best of our knowledge, this is the first method to explicitly address the coupling gain feasibility problem in discrete-time dynamic consensus using a convex optimization framework. In particular, we derive new consensusability conditions that can be formulated as a semidefinite program with LMI constraints (Theorem 4), applicable to both undirected and directed communication graphs.
(ii) The precise source of non-convexity in the consensus problem is identified, and an iterative algorithm (Theorem~5) is developed that provably converges to a stationary point of the original non-convex formulation, thereby substantially improving the feasibility of the consensus problem. This study is the first to provide a formal solution framework for addressing the original non-convex problem.
(iii) The framework is adaptable to
a model-free setup by utilizing efficient input-state data collection schemes (Theorem 6). This data-driven SDP formulation achieves an
equivalent performance to the model-based approach without
introducing conservatism.
(iv) A formal feasibility analysis is provided (Theorem~7) to characterize how the feasibility region of the dynamic consensus problem evolves with the design parameter. (v) A customized algorithm is proposed to balance feasibility, convergence rate,  and energy efficiency according to specific design objectives.

\textbf{Notation:} Let $\mathbb{R}^n$ denote the $n$-dimensional Euclidean space and $\mathbb{R}^{n \times m}$ the set of real $n \times m$ matrices. The sets of symmetric, symmetric positive semidefinite, and symmetric positive definite $n \times n$ matrices are denoted by $\mathbb{S}^n$, $\mathbb{S}_{+}^n$, and $\mathbb{S}_{++}^n$, respectively, with $I_n$ representing the $n \times n$ identity matrix. For a symmetric matrix $X$, the notations $X \succ 0$, $X \prec 0$, $X \succeq 0$, and $X \preceq 0$ indicate positive definite, negative definite, positive semidefinite, and negative semidefinite matrices, respectively. The Kronecker product is denoted by $\otimes$.

 \section{Problem Formulation}\label{sec:3}

Consider a homogeneous linear discrete-time multi-agent system with \( N \) nodes organized over a directed graph \( {\mathcal{G} = (\mathcal{V}, \mathcal{E})} \) with a finite set of \( N \) nodes \( \mathcal{V} = \{ v_1, v_2, \dots, v_N \} \), and a set of directed edges \( \mathcal{E} \subseteq \mathcal{V} \times \mathcal{V} \).
Each directed edge from node \( j \) to node \( i \) is represented as \( (v_j, v_i) \), indicating that information flows from node \( j \) to node \( i \).  The associated adjacency matrix to the graph $\mathcal{G}$ is given as \( A = [a_{ij}] \in \mathbb{R}^{N \times N} \), where the weight \( a_{ij} \) of the edge \( (v_j, v_i) \) is positive if \( (v_j, v_i) \in \mathcal{E} \), otherwise, \( a_{ij} = 0 \). 
 If \( (v_j, v_i) \in \mathcal{E} \), then node \( j \) is considered a neighbor of node \( i \), and the set of neighbors of node \( i \) is denoted by \( \mathcal{A}_i = \{ j \mid (v_j, v_i) \in \mathcal{E} \} \).\\
Define the in-degree matrix \( D = \operatorname{diag}(d_i) \in \mathbb{R}^{N \times N} \), where each \( d_i = \sum_{j \in \mathcal{A}_i} a_{ij} \), and the Laplacian matrix \({\mathcal{L}= D - A} \). Consequently, \(\mathcal{L}\mathbf{1}_N = 0 \), where \( \mathbf{1}_N \) is the \( N \)-dimensional vector of ones. The graph \( \mathcal{G} \) is said to contain a spanning tree if there exists at least one node such that a directed path connects it to every other node in \( \mathcal{G} \). This condition is essential for achieving consensus in multi-agent systems. It implies that \( \lambda_1(\mathcal{L}) = 0 \) is a simple eigenvalue of the associated Laplacian matrix \( \mathcal{L} \), while the other eigenvalues \( \lambda_2(\mathcal{L}), \ldots, \lambda_N(\mathcal{L}) \) have positive real parts \cite{west.2001}.

Let the dynamics of each node be given by
\begin{equation}\label{eq:xi}
   x_i(k + 1) = A x_i(k) + B u_i(k), \quad \forall i \in N, 
\end{equation}
where \( A \in  \mathbb{R}^{n\times n} \) and \( B \in  \mathbb{R}^{n\times m} \).
\begin{assumption}\label{assumption:1}
    The pair \( (A, B) \) is assumed to be controllable, and the directed graph \( \mathcal{G} \) contains a spanning tree. 
\end{assumption}

The multi-agent system \eqref{eq:xi} is given in compact form  by
\begin{equation}\label{eq:compact}
    x(k + 1) = (I_N \otimes A) x(k) + (I_N \otimes B) u(k),
\end{equation}
where \( x(k) = \begin{bmatrix} x_1^T(k) & x_2^T(k) & \cdots & x_N^T(k) \end{bmatrix}^T \in \mathbb{R}^{nN} \) and \( u(k) = \begin{bmatrix} u_1^T(k) & u_2^T(k) & \cdots & u_N^T(k) \end{bmatrix}^T \in \mathbb{R}^{m N} \) represent the global state vector and the global control input, respectively.

Consider implementing the following linear static-feedback control law:
\begin{equation}\label{eq:ui}
    u_i(k) = -c K \sum_{j \in \mathcal{A}_i} a_{ij} \left( x_i(k) - x_j(k) \right),
\end{equation}
where \( c > 0 \) is the coupling gain, and \( K \in \mathbb{R}^{m \times n} \) is the feedback control gain matrix. Then the control signal in a compact form is given by $u(k) = -c (\mathcal{L} \otimes K) x(k)$,
resulting in the overall closed-loop system,
\begin{equation}\label{eq:closed}
    x(k + 1) = \left[ I_N \otimes A - c \mathcal{L} \otimes (B K) \right] x(k).
\end{equation}

The primary objective is to design linear distributed controllers that achieve state synchronization among all the agents. Specifically, this means ensuring that,
\[
\| x_i(k) - x_j(k) \| \to 0 \quad \text{as} \quad k \to +\infty, \quad \forall i, j \in N, \; i \neq j,
\]
which is known as dynamic consensus. A phenomenon in which the interaction among systems gives rise to an emergent behavior that can be interpreted as the weighted average dynamics of the interconnected systems. Dynamic consensus refers to the process where the behavior of each system in a network asymptotically converges over time to align with the collective dynamics of the group. For instance, in a network of oscillators, this results in the group exhibiting dynamics equivalent to those of a single weighted average oscillator. Agreement on potentially unstable trajectories can also be useful. In multi-agent systems, unstable dynamics are sometimes deliberately allowed, either through unstable leader trajectories \cite{ong2021consensus,sun2024adaptive} or the coordinated behavior of inherently unstable agents \cite{feng2021consensusability}. Such trajectories commonly arise in applications like formation control, where agent positions may grow over time, or in systems designed for enhanced maneuverability. If a bounded collective trajectory is required, the standard approach is to stabilize each agent with local feedback before applying the consensus protocol, which is always feasible under the controllability assumption.

\section{Preliminaries}
This section presents the preliminaries for the proposed dynamic consensus control. In this paper, the consensus objective is reformulated as a non-convex problem, where the parameters of the state–action value function in the LQR framework, i.e., the Q-function, serve as the main decision variables. To clarify this connection, we first review the definition and properties of the Q-function in the LQR setting, along with relevant literature results. We then discuss existing consensus conditions for discrete-time multi-agent systems.

Consider the standard infinite-horizon LQR cost
    \begin{equation}\label{eq:J}
        J = \sum_{k=0}^{\infty}  x(k)^T Q x(k) + u(k)^T R u(k) ,
    \end{equation}
    where \( Q \in \mathbb{R}^{n \times n} \) and \( R \in \mathbb{R}^{m \times m} \) are positive semi-definite and positive definite weighting matrices, respectively.
The cost function  at time $k$ for the standard LQR problem is given as
\begin{equation}
V_K\paren{x(k)}=\sum_{i=k}^{\infty}r\paren{x(i),u(i)},
\end{equation}
with $ r\paren{x(i),u(i)}=x(i)^TQx(i) + u(i)^TRu(i)$.
The optimal value function of this problem is known to be quadratic and representable as $ V^*_K\paren{x(k)}=x^T(k)P^*x(k)$ with the optimal control law
\begin{equation}\label{optpolicy}
    u^*(k)=-K^*x(k),
\end{equation}
where $P^*\in \mathbb{S}^n_{++}$ and $K^*\in \mathbb{R}^{m\times n}$. The matrix \( P^* \) satisfies the discrete-time algebraic Riccati equation (DARE) given by \eqref{DARE}, that can be solved efficiently, using the known system matrices \( A \) and \( B \) (\cite{lewis2009})
\begin{equation} \label{DARE}
    P=Q+A^TPA-\paren{A^TPB}\paren{R+B^TPB}^{-1}\paren{B^TPA}.
\end{equation}

Q-learning offers a model-free approach to the LQR problem. With this regard, the Q-function is defined as (\cite{lewis2009})
\begin{equation} \label{Q-function}
    \mathcal{Q}_K\paren{x(k),u(k)} = r\paren{x(k),u(k)} + V_K\paren{x(k+1)},
\end{equation}
which represents the value of taking action $u(k)$ from state $x(k)$ and following the policy $u(k)=-Kx(k)$ afterwards. For the case of LQR problem, the Q-function can  be expressed in quadratic form in both $x(k)$ and $u(k)$  (\cite{lewis2009})

\begin{equation} \label{QfunctionwithH}
 \mathcal{Q}_K\paren{x(k),u(k)}= \begin{bmatrix}
x(k) \\ u(k)
\end{bmatrix}^T H \begin{bmatrix}
x(k) \\ u(k)
\end{bmatrix},
\end{equation}
\begin{equation}
    H = \begin{bmatrix}
Q+A^TPA & \quad A^TPB \\
B^TPA & \quad R+B^TPB
\end{bmatrix}.
\end{equation}
Using the Kronecker product, we have\\
\begin{equation}
 \mathcal{Q}_K\paren{x(k),u(k)}= vec(H)^T \left(\begin{bmatrix}
    x(k)\\
    u(k)
    \end{bmatrix}\otimes\begin{bmatrix}
    x(k)\\
    u(k)
    \end{bmatrix}
    \right),
\end{equation}
where  $vec(H)$ denotes the vector obtained by stacking the columns of the matrix $H$. The optimal Q-function $ \mathcal{Q}^*_K(x(k), u(k))$ represents the cost associated with applying the control input $u(k)$ at time $k$, followed by adhering to the optimal policy $K^*$ in all subsequent steps, i.e.
\begin{align*}
\mathcal{Q}^*_K\paren{x(k),u(k)} & = r(x(k),u(k))+V^*_K(x(k+1))\\
& =\begin{bmatrix}
x(k) \\ u(k)
\end{bmatrix}^T H^* \begin{bmatrix}
x(k) \\ u(k)
\end{bmatrix},
\end{align*}
and the optimal control policy (\ref{optpolicy}) is given by
\begin{align*}
    u^*(k) & =\argmin_u  \mathcal{Q}^*_K\paren{x(k),u(k)}\\
    &=-\paren{R+B^TPB}^{-1}B^TPAx(k),
\end{align*}
with the optimal state feedback gain
\begin{equation*}
    K^* =\paren{R+B^TPB}^{-1}B^TPA.
\end{equation*}

\begin{theorem} \label{thm: bertsekas}(\cite{bertsekas2012dynamic})
   Consider the Riccati equation
   \begin{equation*}
       P_{k+1} = A^T \left( P_k - P_k B (B^T P_k B + R)^{-1} B^T P_k \right) A + Q,
   \end{equation*}
with $ k = 0, 1, \ldots,$ where the initial matrix \( P_0 \) is an arbitrary positive semidefinite symmetric matrix. Assume that the pair \( (A, B) \) is controllable and \( Q =  C^T C \), where the pair \( (A, C) \) is observable. Then, there exists  \( P \in \mathbb{S}^n_{++}\) such that for every \( P_0  \in \mathbb{S}^n_{+} \),  $\lim_{k \to \infty} P_k = P$ and  \( P \) is the unique solution of the algebraic matrix equation
\[
P = A^T \left( P - P B (B^T P B + R)^{-1} B^T P \right) A + Q,
\]
within the class of positive semi-definite matrices.
\end{theorem}

The following theorem presents a semi-definite program whose optimal solution \( H^* \) precisely constructs the optimal value function \(  \mathcal{Q}_K^* \) associated with the objective function \eqref{eq:J}.

\begin{theorem}(\cite{farjadnasab2022model})\label{thm:farjadnasab}
The optimal value function $H^*$ is the solution of the semi-definite program \eqref{eq:farjad} with variables $H=\begin{bmatrix}
    H_{11} & H_{12}\\
    H_{12}^T & H_{22}
\end{bmatrix}\in\mathbb{S}^{n+m}$ and $W\in\mathbb{S}^{n\times n}$, where $\Lambda:=\begin{bmatrix}
Q & 0 \\ 0 & R
\end{bmatrix} \in \mathbb{S}^{(n+m)}_+$ includes the state and input weighting matrices $Q\in \mathbb{S}^{n}_{+}$ and $R\in \mathbb{S}^{m}_{++}$, respectively,

\begin{align}\label{eq:farjad}
   & \maximize_{H,W}~\text{trace}(W)\\
 \nonumber &	\text {subject to~~} \begin{bmatrix}
	H_{11}-W& H_{12} \\
	H_{12}^T & H_{22}
	\end{bmatrix} \succeq 0, \\
\nonumber &	
	\begin{bmatrix}
	&\begin{bmatrix} A \quad B \end{bmatrix}^T H_{11} \begin{bmatrix} A \quad B \end{bmatrix} - H + \Lambda & \begin{bmatrix} A \quad B \end{bmatrix}^T H_{12} \\
	&H_{12}^T \begin{bmatrix} A \quad B \end{bmatrix} & H_{22}
	\end{bmatrix} \succeq 0.
\end{align}
\end{theorem}

Next, Lemma \ref{Lemma:basic} provides necessary and sufficient condition for achieving dynamic consensus in the multi-agent system described by \eqref{eq:closed}.

\begin{lemma}\label{Lemma:basic}
 (\cite{hengster2013synchronization}) Under Assumption \ref{assumption:1}, the multi-agent system \eqref{eq:compact} will reach state consensus under the control law
    \begin{equation}\label{eq:original}
    u_i(k) = -c K \sum_{j \in \mathcal{A}_i} a_{ij} \left( x_i(k) - x_j(k) \right),
\end{equation}
if and only if all matrices \( A - c \lambda_i(\mathcal{L}) B K \) for \( i = 2, \dots, N \) are Schur stable.
\end{lemma}
It should be noted that since the communication graph includes a spanning tree, \( \lambda_1(\mathcal{L}) = 0 \) is a simple eigenvalue of \( \mathcal{L} \), and all the other eigenvalues \( \lambda_2(\mathcal{L}), \dots, \lambda_N(\mathcal{L}) \) in Lemma \ref{Lemma:basic} have positive real parts.
Lemma \ref{Lemma:basic} shows that achieving consensus requires constraints tied to the graph’s spectral properties, complicating the coupling gain design. The following theorem provides sufficient conditions for the existence of the coupling gain, assuming $R= \gamma I$ in standard Riccati equation with $\gamma>0$ as a design parameter.
\begin{theorem}\label{thm:2022}
(\cite{feng2021consensusability})
Under Assumption~\ref{assumption:1} and the additional condition that the matrix \( B \) has full column rank, the multi-agent system  \eqref{eq:compact} achieves consensus when the control law \eqref{eq:ui} is applied with the specified feedback gain
    \begin{equation}
        K = \left( \gamma I_m + B^T P B \right)^{-1} B^T P A,
    \end{equation}
where $P \succ 0$ is the unique positive definite solution of the following discrete-time algebraic Riccati equation
\[
P - A^T P A + A^T P B \left( \gamma I_m + B^T P B \right)^{-1} B^T P A = Q,
\]
if there exists a constant \( c > 0 \) such that all nonzero eigenvalues of the associated Laplacian matrix \( \mathcal{L} \) satisfy 
\begin{equation}
    \min_c \max_{i = 2, \dots, N} \left| 1 - c \lambda_i(\mathcal{L}) \right| < \theta,
\end{equation}
\[
\theta = \sqrt{\frac{\gamma}{\gamma + \lambda_{\max} \left(B^T P B\right)}}.
\]
\end{theorem}

The model-based design in \cite{feng2021consensusability} shows that as $\gamma \to +\infty$, the bound $\theta$ increases toward a critical supremum $\bar{\theta}$. However, large $\gamma$ can cause poor performance, slow convergence, and numerical issues, while small $\gamma$ can render the eigenvalue-dependent constraints infeasible; in the extreme case $\gamma = 0$, $\theta = 0$ and finding a feasible $c$ becomes impossible. Ensuring consensus becomes increasingly difficult as the number of agents grows, especially for Laplacian spectra with complex conjugate or near-zero eigenvalues.

\section{Model-based Dynamic Consensus Analysis and Design}\label{sec:model_based}

In this Section, a new model-based approach is proposed to provide dynamic consensus.  The proposed approach tries to recap the consensus problem by the elements of the Q-functions in LQR. The Q-function is used analytically to represent the consensus problem within a convex optimization framework, and not as a learned quantity. The proposed method relies on solving a set of LMIs for designing the gains, guarantees a pre-specified convergence rate for consensus, and introduces a convex-concave mechanism to enhance the feasibility of the underlying problem.

Consider the Jordan decomposition of the Laplacian matrix \( \mathcal{L} \) as $\mathcal{L} = V J V^{-1},$
where \( V \in \mathbb{R}^{N \times N} \) is a matrix of generalized eigenvectors and \( J \in \mathbb{C}^{N \times N} \) is the Jordan canonical form of \( \mathcal{L} \), consisting of (possibly non-diagonal) Jordan blocks associated with the eigenvalues \( \lambda_i(\mathcal{L}) \). Define the transformed state in modal coordinates as $z(k) = (V^{-1} \otimes I_n) x(k)$.
The dynamics in this coordinate system becomes
\[
z(k+1) = \big(I_N \otimes A - c (J \otimes BK)\big) z(k).
\]
Throughout the remainder of the paper, we denote the eigenvalues of \( \mathcal{L} \) as \( \lambda_i \), with \( \lambda_1 = 0 \) corresponding to the consensus mode. The transformed system has an upper block-triangular structure due to the Jordan form, i.e.
\[
\begin{aligned}
&z(k+1) = 
 \\
&\begin{bmatrix}
A - c \lambda_1 BK & \ast & \cdots & \ast \\
0 & A - c \lambda_2 BK & \ddots & \vdots \\
\vdots & \ddots & \ddots & \ast \\
0 & \cdots & 0 & A - c \lambda_N BK
\end{bmatrix}
z(k),
\end{aligned}
\]

Despite the coupling, the triangular structure ensures that stability of the entire system is determined by the stability of each diagonal block \( A - c \lambda_i BK \). Therefore, dynamic consensus is achieved if all eigenvalues of \( A - c \lambda_i BK \), for \( i = 2, \dots, N \), lie strictly within the unit circle. The consensus rate is dictated by the slowest-decaying disagreement mode, i.e., the largest spectral radius among these blocks. To guarantee a desired convergence level, we require
\[
\rho(A - c \lambda_i BK) < \mu^{-1}, \quad  i = 2, \dots, N,
\]
where \( \mu \geq 1 \) denotes the desired consensus rate.

The following theorem provides new sufficient conditions for consensus with a guaranteed convergence rate. The control and coupling design is formulated as a semidefinite program with LMI constraints, solved locally by each agent in the network.

\begin{theorem}\label{thm:Hbeta}
Let $H^* \in \mathbb{S}^{n+m}$ be the solution of the semi-definite program \eqref{eq:Hmu} with variables $H\in \mathbb{S}^{n+m}$ and ${W \in \mathbb{R}^{n \times n}}$, given by,
\begin{align}\label{eq:Hmu}
  &\maximize_{H,W}~trace(W)\\
 \nonumber &\text{subject to} \begin{bmatrix}
	H_{11}-W& H_{12} \\
	H_{12}^T & H_{22}
	\end{bmatrix} \succeq 0, \\
    \nonumber& \mathcal{M}(H,\mu,A,B) \succeq 0,\quad\text{where}\\
\nonumber&	\mathcal{M}(H, \mu,A,B)=
\end{align}
{\small
\[
\begin{bmatrix}
\mu^2 [A\ B]^T H_{11} [A\ B] - H + \begin{bmatrix} Q & 0 \\ 0 & \gamma I_m \end{bmatrix} & \mu [A\ B]^T H_{12} \\
\mu H_{12}^T [A\ B] & H_{22}
\end{bmatrix}.
\]
}

Next, let $\beta^*$ and $\hat{c}^*$ be the solution of the SDP \eqref{eq:Hbeta} derived by each agent, with variables $\beta \in \mathbb{R}$ and $\hat{c} \in \mathbb{R}$ for $H=H^*$:

{\small
\begin{align}\label{eq:Hbeta}
     &  \maximize_{\beta,\hat{c}}  \beta \\
    \nonumber &\text{subject to}  \begin{bmatrix}
        Q & \beta H_{12}\\
        \beta H_{12}^T & H_{22} 
    \end{bmatrix}\succeq 0,\\
   \nonumber & \mathcal{N}_i(\beta, \hat{c},\lambda_i) \succeq 0, \qquad i=2, \ldots, N\\
   \nonumber & \mathcal{N}_i(\beta, \hat{c},\lambda_i)= \begin{bmatrix}
        \beta -\sqrt{1-\dfrac{\mathcal{R}e{\lambda_i}^2}{|\lambda_i|^2}} & \dfrac{1}{|\lambda_i|}(\hat{c}|\lambda_i|^2-\mathcal{R}e{\lambda_i})\\
       \dfrac{1}{|\lambda_i|}(\hat{c}|\lambda_i|^2-\mathcal{R}e{\lambda_i}) &  \beta +\sqrt{1-\dfrac{\mathcal{R}e{\lambda_i}^2}{|\lambda_i|^2}} 
    \end{bmatrix}.
\end{align}
}

\noindent Then, the distributed control law,
\begin{small}
    \[
\begin{aligned}
    &  u_i(k) = -c^* K^* \sum_{j \in \mathcal{A}_i} a_{ij} \left( x_i(k) - x_j(k) \right),\\
  & c^* = \frac{c_1^* + c_2^*}{2},\qquad K^* = {H_{22}^{*-1}} H_{12}^{*T},
\end{aligned}
\]
\end{small}
\noindent ensures consensus with consensus rate \( \mu \), where,
\begin{small}
    \begin{align*}
    &   c^*_1= \max_i\dfrac{\mathcal{R}e{\lambda_i}-\sqrt{\mathcal{R}e{\lambda_i}^2-|\lambda_i|^2(1-{\beta^*}^2)}}{|\lambda_i|^2},\\
    &    c^*_2 = \min_j\dfrac{\mathcal{R}e{\lambda_j}+\sqrt{\mathcal{R}e{\lambda_j}^2-|\lambda_j|^2(1-{\beta^*}^2)}}{|\lambda_j|^2}.
\end{align*}
\end{small}
\end{theorem}

\textbf{Proof}:
The proof begins by reformulating the consensus condition \eqref{eq:original} in terms of the Q-function elements of a synthetic LQR problem. In the second step, auxiliary variables are introduced to handle the non-convexity of the resulting consensus condition, providing a tractable but conservative representation. Finally, the eigenvalue-dependent constraints are transformed into a set of LMIs, yielding a convex characterization of the feasible coupling gains.\\

Consensus at rate $\mu$ is equivalent to the existence of $P \succ 0$ satisfying $P-\left(\mu A-\mu c\lambda_i B K\right)^*P\left(\mu A-\mu c\lambda_i B K\right) \succ 0$
for $ i=2, \ldots N.$ Assuming $H_{11}, H_{12}, H_{22}$ is the optimal solution of Problem~\eqref{eq:Hmu}, Theorem~\ref{thm:farjadnasab} shows that these correspond to the Q-function subblocks of the LQR problem with dynamics $(\mu A, \mu B)$ and performance matrices $Q$ and $R = \gamma I$. Specifically, $H_{12}=\mu^2 A^TPB$, $H_{22}=\gamma I_m+\mu^2B^TPB$, where $P$ is the solution of the Riccati equation
\begin{equation*}
     P=Q+\mu^2A^TPA-\mu^4\paren{A^TPB}\paren{\gamma I+\mu^2 B^TPB}^{-1}\paren{B^TPA}.
\end{equation*}

Substituting the expressions by their Q-function interpretation as given by $H_{11}, H_{12}, H_{22}$, and also expressing $P$ by the the corresponding Riccati equation results in

\begin{small}
\begin{align}\label{eq:proof1}
 \nonumber   &  P-\left(\mu A-\mu c\lambda_i B K\right)^*P\left(\mu A-\mu c\lambda_i B K\right)\\
\nonumber     &=P-\mu^2A^TPA\\
\nonumber &+c\mu^2\mathcal{R}e{\lambda_i}\left(A^TPBH_{22}^{-1}H_{12}^T+H_{12}H_{22}^{-1}B^TPA\right)\\
\nonumber &-c^2\mu^2(\mathcal{R}e{\lambda_i}^2+\mathcal{I}m{\lambda_i}^2)H_{12}H_{22}^{-1}B^TPBH_{22}^{-1}H_{12}^T\\
\nonumber &=Q- H_{12}H_{22}^{-1}H_{12}^T+
2c\mathcal{R}e{\lambda_i}\left(H_{12}H_{22}^{-1}H_{12}^T\right)\\
\nonumber&-c^2(\mathcal{R}e{\lambda_i}^2+\mathcal{I}m{\lambda_i}^2)H_{12}H_{22}^{-1}(H_{22}-\gamma I_m)H_{22}^{-1}H_{12}^T\\
\nonumber&\succeq Q-\left( I -2c \mathcal{R}e{\lambda_i}+c^2(\mathcal{R}e{\lambda_i}^2+\mathcal{I}m{\lambda_i}^2)\right) H_{12}H_{22}^{-1}H_{12}^T\\
&\succeq Q-\beta^2 H_{12}H_{22}^{-1}H_{12}^T,
\end{align}   
\end{small}
where the auxiliary variable $\beta$ satisfies
\begin{align}\label{eq:beta1}
      & 1  -2c \mathcal{R}e{\lambda_i}+c^2(\mathcal{R}e{\lambda_i}^2+\mathcal{I}m{\lambda_i}^2) < \beta^2,
\end{align}
for $i=2, \ldots N$. From Q-function properties, it is known that $H_{22} \succ 0$. Accordingly, the consensus is guaranteed with the rate $\mu$ if
\begin{align}
    &  \nonumber  \begin{bmatrix}
        Q & \beta H_{12}\\
        \beta H_{12}^T & H_{22} 
    \end{bmatrix}\succeq 0,\\
    \nonumber & 1  -2c \mathcal{R}e{\lambda_i}+c^2(\mathcal{R}e{\lambda_i}^2+\mathcal{I}m{\lambda_i}^2) < \beta^2 ,\\
    \nonumber & i=2, \ldots N.
\end{align}
The inequality \eqref{eq:beta1} defines a set of quadratic constraints on the coupling gain $c$ for $i=2, \ldots N$. It admits a real solution if and only if the discriminants are positive, i.e.,

\begin{align}\label{eq:delta}
\mathcal{R}e\lambda_i^2 - \big(\mathcal{R}e\lambda_i^2 + \mathcal{I}m\lambda_i^2\big) (1 - \beta^2) > 0.
\end{align}
This condition ensures that for each eigenvalue 
$\lambda_i$ the admissible region for $c$ is real. Condition \eqref{eq:delta} can be equivalently expressed as the following lower bound on $\beta$:
\begin{align}\label{eq:betalinear_refined}
\beta > \sqrt{1 - \frac{\mathcal{R}e\lambda_i^2}{|\lambda_i|^2}}, \quad i = 2,\ldots,N.
\end{align}
To guarantee dynamic consensus, there must exist a common value $c$ that satisfies \eqref{eq:beta1} for every 
$i = 2,\ldots,N$. The necessary and sufficient condition for such a $c$ to exist is the nonempty intersection of the feasible intervals determined by each eigenvalue. This leads to the interval consistency condition
\begin{align}\label{eq:minmax_refined}
\nonumber
& \max_i \frac{\mathcal{R}e\lambda_i - \sqrt{\mathcal{R}e\lambda_i^2 - |\lambda_i|^2(1-\beta^2)}}{|\lambda_i|^2}\\
&<
\min_j \frac{\mathcal{R}e\lambda_j + \sqrt{\mathcal{R}e\lambda_j^2 - |\lambda_j|^2(1-\beta^2)}}{|\lambda_j|^2}.
\end{align}
If the inequality \eqref{eq:minmax_refined} holds, then a feasible coupling gain $c$ exists within the admissible interval, satisfying
\begin{align}\label{eq:t1_refined}
\frac{\mathcal{R}e\lambda_i - \sqrt{\mathcal{R}e\lambda_i^2 - |\lambda_i|^2(1-\beta^2)}}{|\lambda_i|^2} < \hat{c},
\end{align}
\begin{align}\label{eq:t2_refined}
\hat{c} < \frac{\mathcal{R}e\lambda_j + \sqrt{\mathcal{R}e\lambda_j^2 - |\lambda_j|^2(1-\beta^2)}}{|\lambda_j|^2}.
\end{align}

Combining \eqref{eq:t1_refined}–\eqref{eq:t2_refined}, the feasibility of $\hat{c}$ and $\beta$ can be expressed equivalently as
\begin{align}\label{eq:beta_general}
\beta^2 \geq \frac{1}{|\lambda_i|^2}(\hat{c}|\lambda_i|^2 - \mathcal{R}e\lambda_i)^2 + 1 - \frac{\mathcal{R}e\lambda_i^2}{|\lambda_i|^2}.
\end{align}
Although the constraint \eqref{eq:beta_general} is generally non-convex, its intersection with the necessary condition \eqref{eq:betalinear_refined} yields a convex admissible set. This is because \eqref{eq:betalinear_refined} permits the use of the Schur complement lemma \cite{boyd2004convex} on \eqref{eq:beta_general}, leading to equivalent LMI conditions $\mathcal{N}_i(\beta, \hat{c}, \lambda_i) \succeq 0$ for $i = 2, \ldots, N$.\\
For any fixed value of $\gamma$, the semidefinite program~\eqref{eq:Hmu} admits an optimal solution $K^\ast$, from which the associated value function is obtained and subsequently used to solve~\eqref{eq:Hbeta}. The feasibility of this optimization problem establishes the existence of a nonempty feasible set for the coupling gain $c$. The midpoint of this interval is then selected as the nominal coupling gain $c^\ast$ for each agent. \qed

Theorem \ref{thm:Hbeta} provides a convex framework for designing distributed control laws implemented by each agent that ensure a consensus rate of $\mu$. However, the convex formulation introduces conservatism due to the first LMI in \eqref{eq:Hbeta}, as detailed in \eqref{eq:proof1}. This conservatism may become significant for large $\gamma$, potentially causing infeasibility. To overcome this limitation, the following theorem presents an iterative convex–concave decomposition algorithm that directly tackles the original non-convex feasibility problem and guarantees convergence to a stationary point.

\begin{theorem}\label{thm:convex-concave}
Let $H \in \mathbb{S}^{n+m}$ be the optimal solution of \eqref{eq:Hmu}. Define the candidate Lyapunov matrix $P = H_{11} - H_{12} H_{22}^{-1} H_{12}^T$, and let $K = H_{22}^{-1} H_{12}^T$. Assume there exists a coupling gain $c>0$ such that the $N-1$ modewise Lyapunov inequalities,
\begin{equation}\label{eq:Lyap_cond_repeat}
 P - \big(\mu A - \mu c \lambda_i B K\big)^T P \big(\mu A - \mu c \lambda_i B K\big)\succ 0,
\end{equation}
are all satisfied for $i=2,\dots,N$. Then Algorithm~\ref{alg:convex_concave} generates a sequence of iterates
$\{(\alpha^{(k)},\beta^{(k)},c^{(k)})\}_{k\ge0}$
such that
\begin{itemize}
    \item[(i)] the scalar objective sequence $\{\alpha^{(k)}\}$ is non-decreasing;
    \item[(ii)] the algorithm (initialized arbitrarily) converges to a limit point that is the stationary point of the original non-convex problem.
\end{itemize}

\end{theorem}

\textbf{Proof}: The proof proceeds in a sequence of steps. First, we restate and reformulate the source of non-convexity in the consensus condition. Next, we show how it can be addressed through a sequence of convexified subproblems, followed by proofs of feasibility preservation, monotonicity, and convergence of the iterative scheme. Finally, we establish certification of consensus for the proposed design.

Step 1- Formulation of the augmented non-convex program: According to the proof of Theorem \ref{thm:Hbeta}, by using the candidate Lyapunov matrix $P = H_{11} - H_{12} {H_{22}}^{-1} H_{12}^T$, the consensus is achieved with consensus rate $\mu$ if
\begin{align*}
    &Q- H_{12}H_{22}^{-1}H_{12}^T+
2c\mathcal{R}e{\lambda_i}\left(H_{12}H_{22}^{-1}H_{12}^T\right)\\
&-c^2(\mathcal{R}e{\lambda_i}^2+\mathcal{I}m{\lambda_i}^2)H_{12}H_{22}^{-1}(H_{22}-\gamma I_m)H_{22}^{-1}H_{12}^T  \\
& = Q-\left( I -2c \mathcal{R}e{\lambda_i}+c^2(\mathcal{R}e{\lambda_i}^2+\mathcal{I}m{\lambda_i}^2)\right) H_{12}H_{22}^{-1}H_{12}^T\\
&+c^2 \gamma (\mathcal{R}e{\lambda_i}^2+\mathcal{I}m{\lambda_i}^2) H_{12}H_{22}^{-2}H_{12}^T \succeq 0,
\end{align*}
which, following similar steps to the proof of Theorem 4, is equivalently described by
\begin{align}\label{eq:Qbetac2}
 \nonumber & {\mathcal{S}_q}_i(\beta, c, \lambda_i, H)=\\
   & \begin{bmatrix}
        Q+c^2 \gamma |\lambda_i|^2 H_{12}H_{22}^{-2}H_{12}^T & \beta H_{12}\\
        \beta H_{12}^T & H_{22} 
    \end{bmatrix}\succeq 0,
\end{align}
and
\begin{align}\label{eq:beta2}
      & 1  -2c \mathcal{R}e{\lambda_i}+c^2(\mathcal{R}e{\lambda_i}^2+\mathcal{I}m{\lambda_i}^2) < \beta^2.
\end{align}
It was previously established that \eqref{eq:beta2} is equivalent to the LMIs $ \mathcal{N}_i(\beta, c,\lambda_i) \succeq 0$ for $i=2, \ldots, N$.
\noindent However, the set of constraints ${\mathcal{S}_q}_i(\beta, c, \lambda_i, H) \succeq 0$ are non-convex in $(\beta,c)$. The objective is to use principles of vectorized convex-concave programming \cite{lipp2016variations} and find a feasible solution for these set of non-convex constraints. To this end, we introduce the augmented non-convex problem
\begin{align}\label{eq:non-convex}
  &  \maximize_{\alpha, \beta, c} \alpha \\
  \nonumber &\text{subject to~~} {\mathcal{S}_q}_i(\beta, c, \lambda_i, H)\succeq \alpha I, \\
\nonumber & \qquad \qquad \quad\mathcal{N}_i(\beta, c,\lambda_i) \succeq 0,  \qquad \qquad   i=2, \ldots, N,
\end{align}
and develop a sequence of convexified subproblems whose solution converges to the stationary point of this problem.

Step 2- Convexified subproblems: The iterative process involves linearizing the \( c^2 \) term in matrix inequalities around the current best estimate of $c$ and solving a convex relaxation of the original non-convex problem at each iteration.
For each fixed iteration $k$, we linearize the scalar function $c^2$ around $c^{(k)}$ by the first-order approximation that preserves the global inequality $2c^{(k)} c - {c^{(k)}}^2\leq c^2$.
Multiplying both sides of this inequality by the positive semidefinite matrix $\gamma|\lambda_i|^2 H_{12}H_{22}^{-2}H_{12}^T\succeq0$ preserves the matrix inequality order, that is
\begin{equation}\label{ineq:quad_linear_bound}
    \left(2c^{(k)}c-{c^{(k)}}^2\right)\gamma|\lambda_i|^2 H_{12}H_{22}^{-2}H_{12}^T
\preceq
c^2\gamma|\lambda_i|^2 H_{12}H_{22}^{-2}H_{12}^T.
\end{equation}

Accordingly, we define the matrix $\mathcal{S}_{l_i}(\beta,c,c^{(k)},\lambda_i,H)$ as
\begin{align*}
  &  \mathcal{S}_{l_i}(\beta,c,c^{(k)},\lambda_i,H)=\\
&\begin{bmatrix}
Q + \left(2c^{(k)}c-{c^{(k)}}^2\right)\gamma|\lambda_i|^2 H_{12}H_{22}^{-2}H_{12}^T & \beta H_{12}\\[4pt]
\beta H_{12}^T & H_{22}
\end{bmatrix}.
\end{align*}
which is affine in the optimization variables $(\alpha,\beta,c)$ and hence the subproblem
\begin{equation}\label{eq:convex_subproblem_repeat}
\begin{aligned}
& \maximize_{\alpha,\beta,c} \ \alpha\\
& \text{subject to}\quad \mathcal{S}_{l_i}(\beta,c,c^{(k)},\lambda_i,H)\succeq \alpha I,\\
&\mathcal N_i(\beta, c,\lambda_i)\succeq0, \qquad i=2,\dots,N,
\end{aligned}
\end{equation}
is a semidefinite program as outlined in Algorithm 1.

Step 3- Feasibility preservation and monotonicity:
First note that an initial feasible point always exists for the subproblem \eqref{eq:convex_subproblem_repeat} due to the unconstrained sign of $\alpha$, i.e., feasibility is trivially guaranteed for sufficiently large negative values of $\alpha$.
Moreover, if $(\alpha^{(k)},\beta^{(k)},c^{(k)})$ satisfies ${\mathcal S}_{q_i}(\beta^{(k)},c^{(k)},\lambda_i,H)\succeq \alpha^{(k)} I$, then, 
$\mathcal{S}_{l_i}(\beta^{(k)},c^{(k)},c^{(k)},H)\succeq \alpha^{(k)} I$, i.e. the same triple is feasible for the convexified problem.
Thus, at each iteration \( k+1 \), the linearization ensures that the feasible region of the newly relaxed problem contains the feasible region of the problem from the previous iteration \( k \). Since the feasible sets progressively expand, and the objective function involves maximizing \( \alpha \), it follows that solving these problems produces a non-decreasing sequence for \( \alpha^{(k)} \). Specifically, if \( \alpha^{(k)} \) represents the maximum value at iteration \( k \), then \( \alpha^{(k+1)} \geq \alpha^{(k)} \), as the feasible region for \( \alpha \) either enlarges or remains unchanged from one iteration to the next.
It remains to show that these set of feasible solutions for the convex subproblem remain feasible for the original problem \eqref{eq:non-convex}. To this end, let $(\alpha^{(k+1)},\beta^{(k+1)},c^{(k+1)})$ denote an optimal solution of the convex subproblem \eqref{eq:convex_subproblem_repeat} and thus
\begin{equation*}
   \mathcal{S}_{l_i}(\beta^{(k+1)},c^{(k+1)},c^{(k)},\lambda_i,H) \succeq \alpha^{(k+1)} I
\end{equation*}
According to \eqref{ineq:quad_linear_bound},
\[
{\mathcal S}_{l_i}(\beta^{(k+1)},c^{(k+1)},c^{(k)},\lambda_i,H)\preceq {\mathcal S}_{q_i}(\beta^{(k+1)},c^{(k+1)},\lambda_i,H).
\]
Thus ${\mathcal S}_{q_i}(\beta^{(k+1)},c^{(k+1)},\lambda_i,H)\succeq \alpha^{(k+1)} I$.  Therefore, the point $(\alpha^{(k+1)},\beta^{(k+1)},c^{(k+1)})$ is also feasible for the original non-convex problem \eqref{eq:non-convex}.

Step 4 — Certification of consensus:
According to the convex–concave framework \cite{lipp2016variations}, any limit point of the sequence $\{(\alpha^{(k)},\beta^{(k)},c^{(k)})\}$ generated by repeated linearization \eqref{eq:convex_subproblem_repeat} in Algorithm 1 is a stationary point of the original non-convex problem \eqref{eq:non-convex}. Assuming that the stationary points of the original problem \eqref{eq:non-convex} satisfy $\alpha>0$, guarantees that the limit point obtained through Algorithm 1 satisfies $\alpha>0$. By the equivalences in Steps 1, $\alpha>0$ implies that for the corresponding $(\beta,c)$ all constraints ${\mathcal S}_{q_i}(\beta,c,\lambda_i,H)\succeq 0$ and $\mathcal N_i(\beta,\hat c,\lambda_i)\succeq0$ hold; hence the consensus inequalities \eqref{eq:Lyap_cond_repeat} hold for all $i=2,\dots,N$. which corresponds to the condition for consensus at rate $\mu$, under the Lyapunov candidate ${P = H_{11} - H_{12} {H_{22}}^{-1} H_{12}^T}$. 
\qed

\begin{algorithm}
\caption{Design of Feasible Coupling Gain for Consensus of Homogeneous Multi-agent Systems}
\label{alg:convex_concave}
\begin{algorithmic}[1]
\Require \( H_{11}, H_{12}, H_{22}, Q, \gamma \), the spectrum of Laplacian $\Lambda$, termination tolerance \( \epsilon \).
\Ensure The coupling gain \( c \).
\State Initialize \( c^{(0)} \), \( s = 0 \), \( k = 0 \), and \( \alpha_{\text{p}} \).
\While{ \( s = 0 \)}
    \State Solve the semi-definite program:
    \begin{align*}
      & \underset{\alpha, \beta, c}{\text{maximize~~}} \alpha \\
     \nonumber &\text{subject to~~~} \mathcal{S}_{l_i}(\beta, c, c^{(k)}, \lambda_i, H) \succeq \alpha I, \\
      & \qquad \qquad \quad \mathcal{N}_i(\beta, c, \lambda_i) \succeq 0, \qquad \qquad i = 2, \dots, N.
    \end{align*}
    \State Update \( c^{(k+1)} = c \), \( \beta^{(k+1)} = \beta \), and \( \alpha^{(k+1)} = \alpha \).
     \If{\( |\alpha^{(k+1)} - \alpha_{\text{p}}| \leq \epsilon \)}
        \State Set \( s = 1 \).
        \If{\( \alpha^{(k+1)} \geq  0 \)}
            \State \textbf{Terminate} the algorithm with the feasible \par solution    \( c^{(k+1)}, \beta^{(k+1)} ,\alpha^{(k+1)} \).
        \Else
            \State \textbf{Terminate} the algorithm and return 
            \(\alpha^{(k+1)} \) \par (Infeasible non-convex constraints).
        \EndIf
    \EndIf
    \State Set \( \alpha_{\text{p}} = \alpha^{(k+1)} \).
    \State Set \( k = k + 1 \).
\EndWhile
\end{algorithmic}
\end{algorithm}

The computational cost per agent in Algorithm 1 mainly depends on the sum of the local state and input dimensions $(n+m)$ rather than on the total number of agents $N$. That is because, although the number of LMI constraints increases linearly with the number of agents, the dominant computational cost in SDPs depends on the number of scalar decision variables and the dimension of the largest LMI block. In our formulation, this block has dimension $n+m$, which is independent of the network size. Consequently, the overall computation time scales favorably with the number of agents.

\begin{remark}
    Although each agent implements the control law locally, the design requires knowledge of the graph Laplacian spectrum, which reflects the network topology. This requires partial global information sharing. As such, the solution is distributed in terms of computation, but it is not fully decentralized in the strictest sense. The approach is applicable when the communication topology is known or can be estimated.
\end{remark}

\section{The Proposed Model-free Consensus Algorithm}

To solve the dynamic consensus problem in Section \ref{sec:model_based}, knowledge of the system dynamics is required. However, in practice, precise knowledge of the system dynamics is often difficult to obtain, or may even be unavailable. In this section, we demonstrate that it is possible to bypass the dependency on system dynamics in the model-based algorithm presented in Section \ref{sec:model_based} by using data from each agent’s inputs and states. It is notable that the model-free extension in this Section is not based on Q-learning, but rather uses the Q-function-inspired reformulation within a data-driven SDP setup.

Define \( D_j \in \mathbb{R}^{(n+m) \times l} \) as the trajectory data matrix for the inputs and states of agent \( j \), for \( j \in \{1, 2, \dots, N\} \)

\begin{equation} \label{Ddef}
D_j:=\begin{bmatrix}
x_j(k_0) & x_j(k_0+1) & \cdots & x_j(k_0+l-1) \\
u_j(k_0) & u_j(k_0+1) & \cdots & u_j(k_0+l-1)
\end{bmatrix},
\end{equation}
where $x_j(k)=[x_{j1}(k)\quad x_{j2}(k)\quad \cdots\quad x_{jn}(k)]^T\in \mathbb{R}^n$, ${u_j(k)=[u_{j1}(k)\quad u_{j2}(k)\quad \cdots\quad u_{jm}(k)]^T\in \mathbb{R}^m}$, and $k_0$ is the index of the first sample to be collected. Define ${X_{Dj} \in \mathbb{R}^{n\times l}}$ as
\begin{equation}\label{Xdef}
\begin{aligned}
X_{Dj}:=&\begin{bmatrix} A \quad B \end{bmatrix}D_j \\
=&\left[\begin{matrix}
Ax_j(k_0)+Bu_j(k_0) \mkern6mu Ax_j(k_0+1)+Bu_j(k_0+1) \end{matrix}\right.\\
&\qquad \left.\begin{matrix} \cdots \mkern6mu Ax_j(k_0+l-1)+Bu(k_0+l-1)
\end{matrix}\right]\\
=&\begin{bmatrix}
x_j(k_0+1) & x_j(k_0+2) & \cdots & x_j(k_0+l)
\end{bmatrix}.
\end{aligned}
\end{equation}

Before introducing the main data-dependent formulation of the dynamic consensus problem, note that satisfying the full-rank condition for $D_j$ requires the input signal $u_j[k_0, k_0+l-1]$ (the restriction of the sequence $u$ to the interval $[k_0, k_0+l-1]$) to be PE of order $L$, with $L \geq n+1$ \cite{willems2005note}. The formal definition of persistent excitation for the signal $u$ over the interval $[k_0, k_0+l-1]$ is provided below.

\begin{definition} \cite{willems2005note}
The signal $u_{[k_0,k_0+l-1]}$ is PE of order $L$, if the matrix $\mathcal{H}_L(u)$ has full row rank,
\small{
\begin{equation}\label{eq:PE}
  \mathcal{H}_L(u)=  \begin{bmatrix}
    u(k_0) & u(k_0+1) & \ldots & u(k_0+l-L)\\
     u(k_0+1) & u(k_0+2) & \ldots & u(k_0+l-L+1)\\
     \vdots & \vdots & \ddots & \vdots \\
      u(k_0+L-1) & u(k_0+L) & \ldots & u(k_0+l-1)\\
    \end{bmatrix}.
\end{equation}
}
\end{definition}
In practice, to satisfy the PE condition, a small probing noise, typically consisting of Gaussian white noise or a combination of sinusoids, needs to be introduced into the control signal.\\

Theorem \ref{thm:Hbetadata} presents semi-definite programs for solving the consensus problem without requiring any knowledge of the system dynamics. The data-dependent counterpart of Theorem \ref{thm:Hbeta} is provided in Theorem \ref{thm:Hbetadata}, which facilitates the derivation of the same distributed control design as in Theorem \ref{thm:Hbeta}, provided that at least \( m + n + 1 \) samples are collected by each agent to form the matrices \( D_j \) and \( X_{Dj} \), and the control signals are PE of order at least \( n+1 \).

\begin{theorem}\label{thm:Hbetadata}
Let $H^* \in \mathbb{S}^{n+m}$ be the solution of the semi-definite program \eqref{eq:Hmudata} with variables $H\in \mathbb{S}^{n+m}$ and $W \in \mathbb{R}^{n \times n}$, given by
\small{
\begin{align}\label{eq:Hmudata}
  &\maximize_{H,W} trace(W)\\
   \nonumber &\text{subject to~~~} \begin{bmatrix}
	H_{11}-W& H_{12} \\
	H_{12}^T & H_{22}
	\end{bmatrix} \succeq 0, \\
    \nonumber& \mathcal{M}_D(H,\mu,D_j,X_{Dj}) \succeq 0,\quad\text{where}\\
\nonumber&	
	\mathcal{M}_D(H,\mu,D_j,X_{Dj})=\\
   \nonumber &	\begin{bmatrix}
	\mu^2 X_{Dj}^TH_{11}X_D-D_j^THD_j+D_j^T\begin{bmatrix}
	    Q & 0\\
        0 & \gamma I
	\end{bmatrix}D_j & \mu X_{Dj}^TH_{12} \\
	\mu H_{12}^TX_{Dj} & H_{22}
	\end{bmatrix}.
\end{align}
}
\normalsize
\noindent where \( D_j \in \mathbb{R}^{(n+m) \times l} \) and \({X_{Dj} \in \mathbb{R}^{n\times l}}\)represents the trajectory data matrices \eqref{Ddef} and \eqref{Xdef} for the inputs and states of agent \( j \), with \( j \in \{1, 2, \dots, N\} \), and the sequence of the input signal $u_j[k_0, k_0+l-1]$ is PE of order at least $n+1$.
Next, let $\beta^*$ and $\hat{c}^*$ be the solution of the SDP \eqref{eq:Hbeta2} with variables $\beta \in \mathbb{R}$, $\hat{c} \in \mathbb{R}$ for $H=H^*$:
\begin{align}\label{eq:Hbeta2}
     &  \maximize_{\beta,\hat{c}}  \beta \\
   \nonumber &\text{subject to~~~} \begin{bmatrix}
        Q & \beta H_{12}\\
        \beta H_{12}^T & H_{22} 
    \end{bmatrix}\succeq 0,\\
   \nonumber &\qquad \qquad \quad \mathcal{N}_i(\beta, \hat{c},\lambda_i) \succeq 0, \qquad i=2, \ldots, N.
\end{align}
Then, the distributed control law,
\[
\begin{aligned}
    &  u_i(k) = -c^* K^* \sum_{j \in \mathcal{A}_i} a_{ij} \left( x_i(k) - x_j(k) \right),\\
  & c^* = \frac{c_1^* + c_2^*}{2},\qquad K^* = {H_{22}^{*-1}} H_{12}^{*T},
\end{aligned}
\]
which is derived by each agent, ensures consensus with a guaranteed consensus rate \( \mu \), where,
\begin{align*}
    &   c^*_1= \max_i\dfrac{\mathcal{R}e{\lambda_i}-\sqrt{\mathcal{R}e{\lambda_i}^2-|\lambda_i|^2(1-{\beta^*}^2)}}{|\lambda_i|^2},\\
    &    c^*_2 = \min_j\dfrac{\mathcal{R}e{\lambda_j}+\sqrt{\mathcal{R}e{\lambda_j}^2-|\lambda_j|^2(1-{\beta^*}^2)}}{|\lambda_j|^2}.
\end{align*}
\end{theorem}

\textbf{Proof}: It suffices to show the equivalence between the data-dependent SDP \eqref{eq:Hmudata} and the model-based SDP \eqref{eq:Hmu} in Theorem \ref{thm:Hbeta}. Note that by Schur complement lemma \cite{boyd2004convex}, the second constraint in model-based SDP \eqref{eq:Hbeta}, i.e., $\mathcal{M}(H,\mu,A,B) \succeq 0$ is equivalently described as

\begin{align}\label{eq:schur}
  & \mu^2 \begin{bmatrix} A \quad B \end{bmatrix}^T(H_{11}-H_{12}H_{22}^{-1}H_{12}^T) \begin{bmatrix} A \quad B \end{bmatrix}\\
  \nonumber &- H + \begin{bmatrix}
       Q & 0\\
       0 & \gamma I_m
   \end{bmatrix}  \succeq0.
\end{align}
Moreover, since the input signal $u_j[k_0, k_0+l-1]$ is PE of order at least $n+1$, the matrix \( D_j \) has a full row rank \cite{willems2005note}. Accordingly, based on the properties of matrix congruence, it becomes possible to apply a multiplication to the inequality constraint \eqref{eq:schur} from left and right by $D_j^T$ and $D_j$, to derive the equivalent constraint

\begin{align}
  &  \mu^2 D_j^T \begin{bmatrix} A & B \end{bmatrix}^T (H_{11} - H_{12} H_{22}^{-1} H_{12}^T) \begin{bmatrix} A & B \end{bmatrix} D_j \nonumber \\
    & - D_j^T \left( H - \begin{bmatrix}
        Q & 0 \\
        0 & \gamma I_m
    \end{bmatrix} \right) D_j \succeq 0 \label{cond3} \\
   & \Rightarrow \mu^2 X_{Dj}^T (H_{11} - H_{12} H_{22}^{-1} H_{12}^T) X_{Dj} \nonumber \\
    & - D_j^T \left( H - \begin{bmatrix}
        Q & 0 \\
        0 & \gamma I_m
    \end{bmatrix} \right) D_j \succeq 0. \label{cond4}
\end{align}

Finally, because $H_{22} \succ 0$, by directly applying the Schur complement, the constraint (\ref{cond4}) is equivalent to
\begin{equation}
\begin{bmatrix}
\mu^2 X_{Dj}^TH_{11}X_{Dj}-D_j^T(H-\Lambda)D_j & \mu X_{Dj}^TH_{12} \\
\mu H_{12}^TX_{Dj} & H_{22}
\end{bmatrix} \succeq 0,
\end{equation}
with  $\Lambda=\begin{bmatrix}
Q & 0 \\ 0 & \gamma I
\end{bmatrix}$, and corresponds to the data-based SDP $\mathcal{M}_D(H,\mu,D_j,X_{Dj}) \succeq 0$.
 \qed

To guarantee that the matrix $D_j$ has full row rank, i.e., a column rank of $m+n$, the number of samples $l$ must be at least equal to the sum of the number of inputs and states for every agent. Therefore, a minimum of $l+1 = m+n+1$ samples is required.

The proposed method is a non-iterative technique that computes the optimal feedback gain for agent $j$ and the coupling gain $c$ in a single step, utilizing \( l+1 = m+n+1 \) samples of inputs and state trajectories. It is also possible to enhance the feasibility of this data-dependent problem for the coupling gain by using the proposed convex-concave decomposition introduced in Algorithm \ref{alg:convex_concave}.\\

\begin{remark}\label{remark:gamma}
Excessive increases or decreases in \(\gamma\) can adversely affect the transient response of dynamic consensus. It is also notable that tuning $Q$ could offer additional flexibility. In our implementation we chose to fix $Q$ for simplicity and explore the solution behavior by adjusting \( \gamma \), which offers a more interpretable and scalar tuning mechanism. Specifically, as \(\gamma \to \infty\), the feedback gain \(K\) approaches a form where \(Q \to 0\). In the absence of a guaranteed consensus rate $\mu$, it would result in significantly slower consensus dynamics. Moreover, it makes the feasibility of coupling gain challenging, even by solving the non-convex constraint in Theorem \ref{thm:convex-concave}. On the other side, excessive decreases in $\gamma$ results in excessive control signals, sensitivity to noise and ill-conditioning of the optimization problems.
\end{remark}

To address the trade-offs discussed in Remark~\ref{remark:gamma}, Theorem~\ref{thm:gamma} characterizes how variations in the tuning parameter~$\gamma$ influence the feasibility of the problem, particularly regarding the existence of a coupling gain.

\begin{theorem}\label{thm:gamma}
Let $(H_{\gamma_1}, W_{\gamma_1})$ and $(H_{\gamma_2}, W_{\gamma_2})$ be the corresponding solutions obtained from solving the data-driven optimization problem \eqref{eq:Hmudata} with parameters $\gamma_1$ and $\gamma_2$, respectively. Assume that the semi-definite program \eqref{eq:Hbeta2} is feasible in both cases yielding the optimal values $\beta_{\gamma1}$ and $\beta_{\gamma2}$. If $\gamma_1 < \gamma_2$, we have $W_{\gamma_1} \preceq W_{\gamma_2}$. Moreover, for strictly unstable matrix $A$, it follows that $\beta_{\gamma1} \geq \beta_{\gamma2}$.
\end{theorem}

\textbf{Proof.}
The proof begins by establishing, via induction, that the optimal value of $W$ increases monotonically with $\gamma$. This result will then be used to show that, for strictly unstable systems, the optimal value of $\beta$ decreases monotonically as $\gamma$ increases.

To formalize the monotonic dependence of $W$ on $\gamma$, define a sequence of matrices \( P_{\gamma_1}^{(k)} \) and \( P_{\gamma_2}^{(k)} \) for two DAREs with \( R = \gamma_1 I \) and \( R = \gamma_2 I \), respectively, such that \( P_{\gamma_1}^{(0)} = P_{\gamma_2}^{(0)} = 0 \),
 and \( P_{\gamma_1}^{(k+1)} \) and \( P_{\gamma_2}^{(k+1)} \) are computed from the previous iteration using the Riccati equation, as given in Theorem \ref{thm: bertsekas}. Suppose that \( P_{\gamma_1}^{(k)} \preceq P_{\gamma_2}^{(k)} \) holds for some \( k > 0 \).
We will show that this implies \( P_{\gamma_1}^{(k+1)} \preceq P_{\gamma_2}^{(k+1)} \).
The DARE updates for each \( P^{(k+1)} \) can be written as
\small{\[
P_{\gamma_1}^{(k+1)} = A^T P_{\gamma_1}^{(k)} A - A^T P_{\gamma_1}^{(k)} B (\gamma_1 I + B^T P_{\gamma_1}^{(k)} B)^{-1} B^T P_{\gamma_1}^{(k)} A + Q
\]
\[
P_{\gamma_2}^{(k+1)} = A^T P_{\gamma_2}^{(k)} A - A^T P_{\gamma_2}^{(k)} B (\gamma_2 I + B^T P_{\gamma_2}^{(k)} B)^{-1} B^T P_{\gamma_2}^{(k)} A + Q
\]}
\normalsize
According to the monotonicity property of the Riccati operator\cite{bitmead1985monotonicity}, $P_{\gamma_1}^{(k)} \preceq P_{\gamma_2}^{(k)}$ gives
\begin{small}
    \begin{equation}\label{eq:gamma1}
    P_{\gamma_1}^{(k+1)} \preceq A^T P_{\gamma_2}^{(k)} A - A^T P_{\gamma_2}^{(k)} B (\gamma_1 I + B^T P_{\gamma_2}^{(k)} B)^{-1} B^T P_{\gamma_2}^{(k)} A + Q.
\end{equation}
\end{small}
Moreover, Since $\gamma_1 < \gamma_2$, we have,
\begin{equation}\label{eq:gamma2}
    (\gamma_1 I + B^T P_{\gamma_1}^{(k)} B )^{-1} \succeq (\gamma_2 I + B^T P_{\gamma_1}^{(k)} B)^{-1}.
\end{equation}
The inequalities \eqref{eq:gamma1} and \eqref{eq:gamma2} together result in
\begin{small}
    \begin{equation*}
    P_{\gamma_1}^{(k+1)} \preceq A^T P_{\gamma_2}^{(k)} A - A^T P_{\gamma_2}^{(k)} B (\gamma_2 I + B^T P_{\gamma_2}^{(k)} B)^{-1} B^T P_{\gamma_2}^{(k)} A + Q,
\end{equation*}
\end{small}

 or equivalently, \( P_{\gamma_1}^{(k+1)} \preceq P_{\gamma_2}^{(k+1)} \).
By induction, this inequality holds for all \( k \geq 0 \).

As \( k \to \infty \), \(  P_{\gamma_1}^{(k)} \to  P_{\gamma_1} \) and \(  P_{\gamma_2}^{(k)} \to  P_{\gamma_2} \), where \(  P_{\gamma_1} \) and \(  P_{\gamma_2} \) are the unique solutions to the DAREs with \( R = \gamma_1 I \) and \( R = \gamma_2 I \), respectively.
Since  \( P_{\gamma_1}^{(k)} \preceq  P_{\gamma_1}^{(k)} \) for all \( k \geq 0 \), it follows that \( P_{\gamma_1} \preceq P_{\gamma_2} \) in the limit. Finally, replacing $W_{\gamma_1}=P_{\gamma_1}$ and $W_{\gamma_2}=P_{\gamma_2}$ according to \cite{farjadnasab2022model} gives, $W_{\gamma_1} \preceq W_{\gamma_2}$.

For the strictly unstable matrix $A$, consider the Lyapunov candidate $\hat{P}=P_{\gamma_2} -P_{\gamma_1} \succeq 0 $. As all the eigenvalues of $A$ are outside of the unit circle, ${ \hat{P}-A^T \hat{P} A \preceq 0,}$
or equivalently,
\begin{equation*}
   ( P_{\gamma_2} -P_{\gamma_1})-A^T ( P_{\gamma_2} -P_{\gamma_1}) A \preceq 0.
\end{equation*}

Replacing
\begin{align*}
   & P_{\gamma_1}=W_{\gamma_1}= {H_{11}}_{\gamma_1}- {H_{12}}_{\gamma_1}{H_{22}}^{-1}_{\gamma_1}{H_{12}}_{\gamma_1}^T, \\
  &  P_{\gamma_2} = W_{\gamma_2}=  {H_{11}}_{\gamma_2}- {H_{12}}_{\gamma_2}{H_{22}}^{-1}_{\gamma_2}{H_{12}}_{\gamma_2}^T,\\
  & {H_{11}}_{\gamma_1}=A^T P_{\gamma_1} A+Q,\\
  & {H_{11}}_{\gamma_2}=A^T P_{\gamma_2} A+Q,
\end{align*}
results in
\begin{equation}\label{eq:H1222}
  {H_{12}}_{\gamma_1}{H_{22}}^{-1}_{\gamma_1}{H_{12}}_{\gamma_1}^T \preceq  {H_{12}}_{\gamma_2}{H_{22}}^{-1}_{\gamma_2}{H_{12}}_{\gamma_2}^T .
\end{equation}
The first LMI in \eqref{eq:Hbeta2}, $Q \succeq \beta^2 H_{12} H_{22}^{-1} H_{12}^T$, depends on the optimal solution of \eqref{eq:Hmudata}, whereas the remaining LMIs in \eqref{eq:Hbeta2} depend only on the Laplacian eigenvalues and are independent of $\gamma$. Hence, given feasibility, \eqref{eq:H1222} implies that the corresponding optimal parameters satisfy $\beta_{\gamma_1} \geq \beta_{\gamma_2}$.
  \qed

As a corollary of Theorem \ref{thm:gamma}, we can conclude that for strictly unstable systems, if the constraints in \eqref{eq:Hbeta2} are infeasible with \( (H_{\gamma_1}, W_{\gamma_1}) \), then these constraints will also be infeasible with \( (H_{\gamma_2}, W_{\gamma_2}) \) where \( \gamma_1 < \gamma_2 \). 
This indicates that increasing \(\gamma\) makes the feasibility conditions progressively more restrictive. In practice, one can start from a small feasible \(\gamma\) and gradually increase it until infeasibility occurs. The largest feasible \(\gamma\) defines the most energy-efficient solution that the given network and agent dynamics can sustain without violating the certified consensus rate. A summary of this procedure incorporated with optimization problem \eqref{eq:Hbeta2} or alternatively the sequence of problems according to Algorithm \ref{alg:convex_concave} is presented in Algorithm \ref{alg:gamma}.

\section{Simulation Results}
\begin{table*}[ht]
\centering
\caption{Feasibility of the consensus procedures for different values of $\gamma$ in Example 3.}
\label{tab:gamma1}
\begin{tabular}{cccccccc}
\hline
\textbf{Method} & $\gamma = 0$ & $\gamma = 0.01$ & $\gamma = 0.1$ & $\gamma = 1$ & $\gamma = 10$ & $\gamma = 100$ & $\gamma = 1000$ \\
\hline
Theorem 6 & \textbf{Feasible} & \textbf{Feasible} & \textbf{Feasible} & Infeasible & Infeasible & Infeasible & Infeasible \\
Algorithm 1 & \textbf{Feasible} & \textbf{Feasible} & \textbf{Feasible} & \textbf{Feasible} & \textbf{Feasible} & \textbf{Feasible} & \textbf{Feasible} \\
\cite{feng2021consensusability,feng2021q} & Infeasible & Infeasible & Infeasible & Infeasible & Feasible & \textbf{Feasible} & \textbf{Feasible} \\
\hline
\end{tabular}
\end{table*}

\begin{algorithm}[H]
\caption{Data-Driven Consensus of Homogeneous Multi-agent Systems}
\label{alg:gamma}
\begin{algorithmic}[1]
\Require Matrix $Q$, consensus rate $\mu$, initial value $\gamma_0$, incremental step $\Delta \gamma$, and maximum allowable value $\gamma_{max}$.
\Ensure The feedback gain $K$ and coupling gain $c$.

\State \textbf{Initialize:} Set \( s = 0, l=0 \), and $\gamma_l =\gamma_0$.
  \State select an initial policy (non necessarily stabilizing) and collect at least $n+m+1$ input and state data samples.
  \State define the trajectory matrices \( D_j \in \mathbb{R}^{(n+m) \times l} \) and \( X_{D_j} \in \mathbb{R}^{(n) \times l} \) using available data, for each agent $j\in \{1, \ldots N\}$.
  \State Ensure \( D_j \) has column rank \( m + n \).
\While{$s=0$ and $\gamma_l \leq \gamma_{max}$}
    \State Solve the data-dependent optimization problem \eqref{eq:Hmudata} with $\gamma_l$ and derive the matrices $H_{11}, H_{12}, H_{22}$.
    \State Solve the optimization problem \eqref{eq:Hbeta2} to find $\beta$ or alternatively solve the sequence of problems according to Algorithm \ref{alg:convex_concave}.
   \If {\eqref{eq:Hbeta2} is feasible or alternatively Algorithm \ref{alg:convex_concave} gives a feasible solution}
        \State $K ={H_{22}^*}^{-1} H_{12}^{*T}$
        \State $c = c^*$
        \Else
        \State \( s = 1 \) \text{ (Infeasibility detected)}
    \EndIf
    \State Update \( l = l + 1 \).
    \State Update \( \gamma_l = \gamma_l + \Delta\gamma \).
\EndWhile
\State \textbf{Output:}   $u_j(k) = -c K \sum_{i \in \mathcal{N}_j} a_{ji} \left( x_j(k) - x_i(k)\right)$ ,
\end{algorithmic}
\end{algorithm}

\begin{figure}[h]
    \centering
    \includegraphics[width=0.6\linewidth]{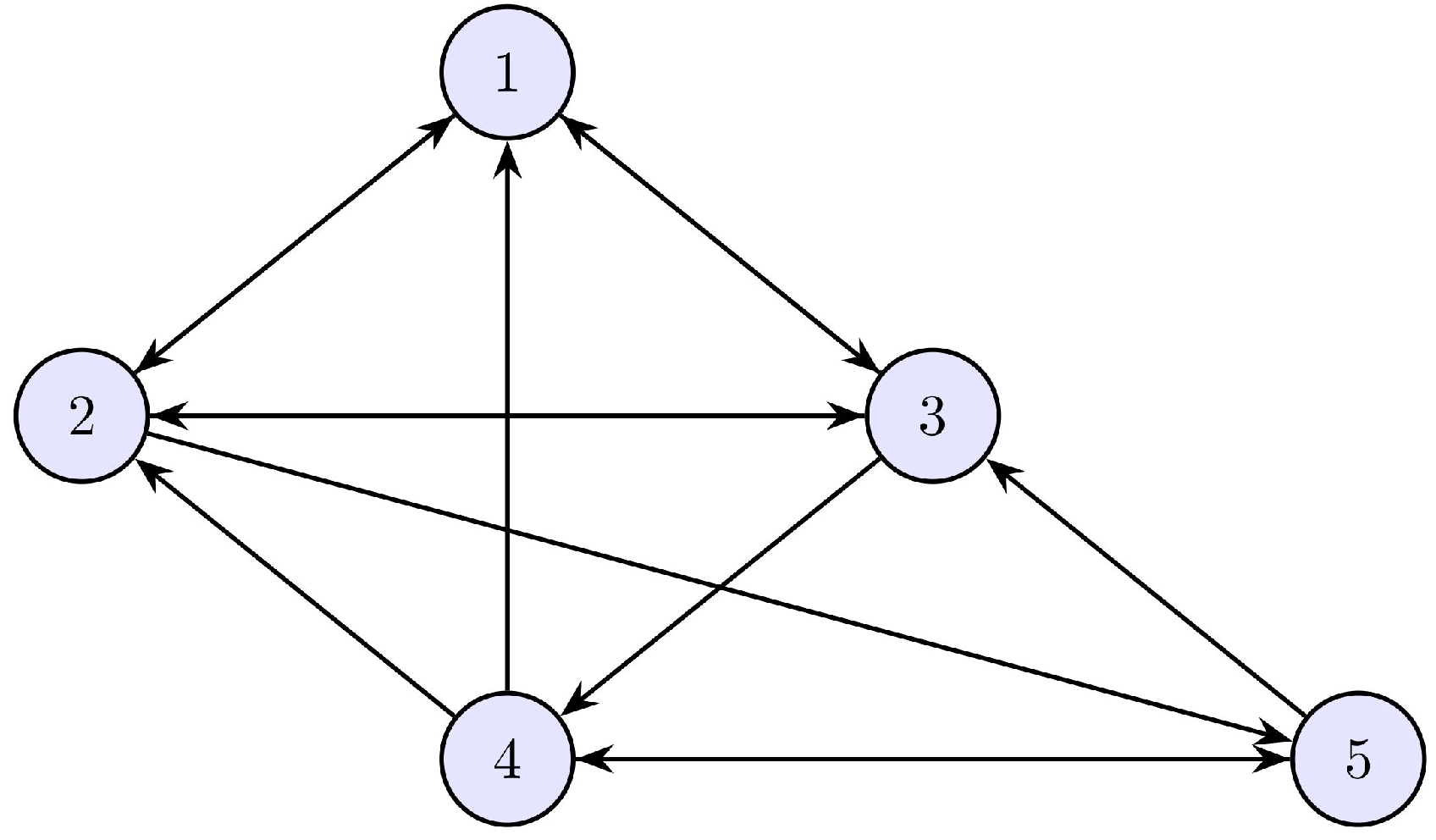}
    \caption{The communication graph in Example 1.}
    \label{fig:GExample1}
\end{figure}
\subsection*{Example 1:}
Consider a multi-agent system with \( N = 5 \) agents, where the dynamics of each agent is given by
\[
A = \begin{bmatrix}
1 & 4 & -0.7 \\
0 & 1 & 0 \\
0 & 0 & 1
\end{bmatrix}, \quad
B = \begin{bmatrix}
0 & 1 \\
2 & 0 \\
0 & 3
\end{bmatrix}.
\]
Let $Q= I$ and $\gamma=100$, and $\mu=1.2$. Consider the connectivity graph $\mathcal{G}$ be given as Figure \ref{fig:GExample1}.  The Laplacian graph associated to $\mathcal{G}$ is
\[
\mathcal{L} = 
\begin{bmatrix}
  3 & -1 & -1 & -1 &  0 \\
 -1 &  3 & -1 & -1 &  0 \\
 -1 & -1 &  3 &  0 & -1 \\
  0 &  0 & -1 &  2 & -1 \\
  0 & -1 &  0 & -1 &  2 
\end{bmatrix}.
\]

\begin{figure}[ht]
    \centering
    \begin{subfigure}[t]{0.5\textwidth}
        \centering
        \includegraphics[width=0.75\textwidth]{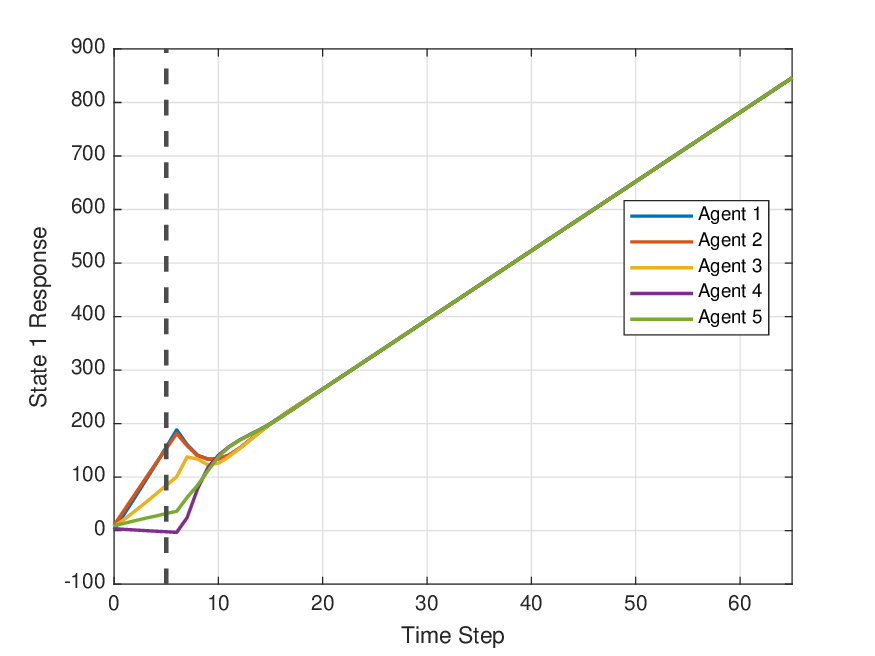}
        \caption{}
        \label{fig:subfig1}
    \end{subfigure}
    
    \begin{subfigure}[t]{0.5\textwidth}
        \centering
        \includegraphics[width=0.75\textwidth]{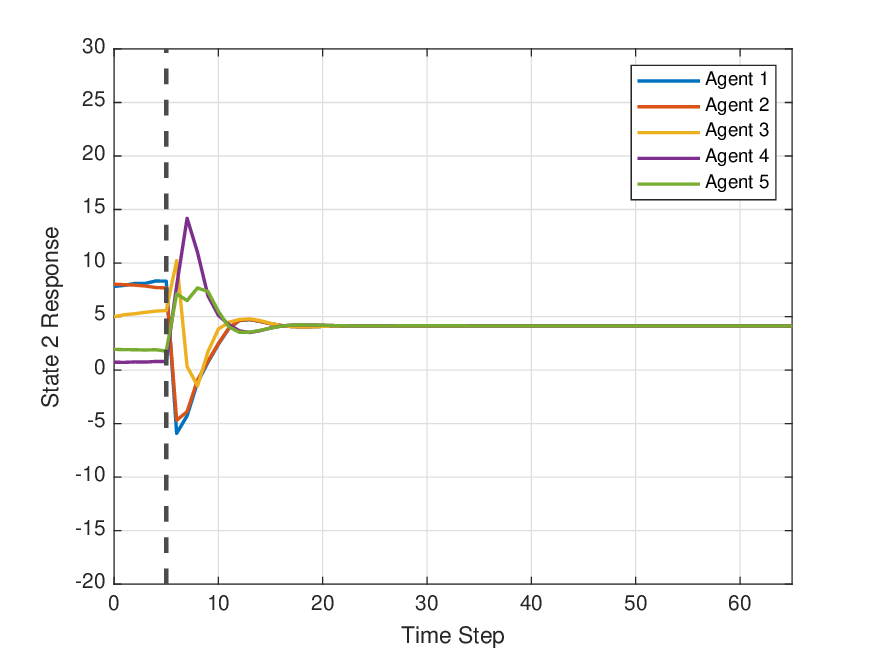}
        \caption{}
        \label{fig:subfig2}
    \end{subfigure}
    
    \begin{subfigure}[t]{0.5\textwidth}
        \centering
        \includegraphics[width=0.75\textwidth]{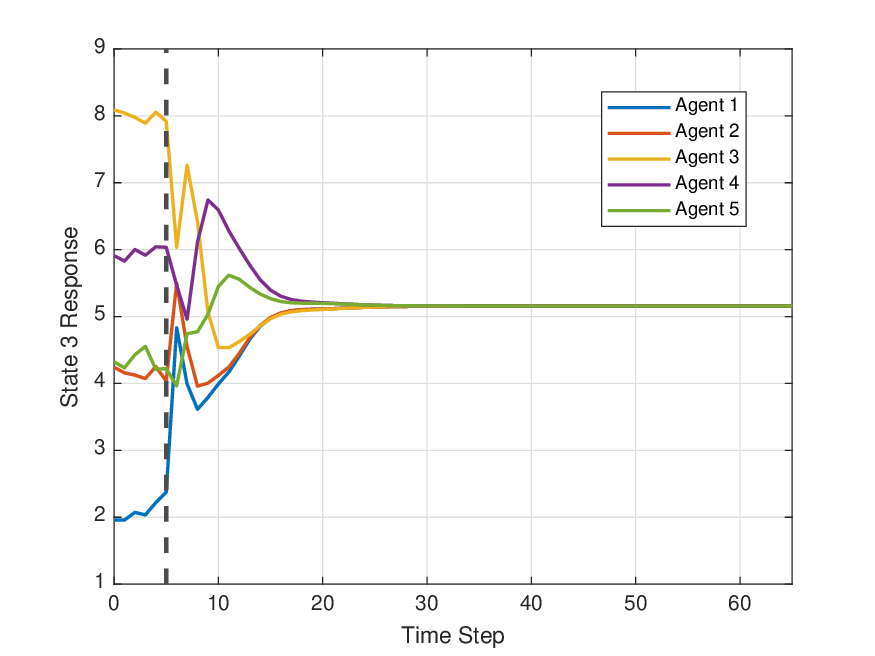}
        \caption{}
        \label{fig:subfig3}
    \end{subfigure}

    \caption{Dynamic consensus in Example 1 with  (a) State $1$ of all the agents, (b) State $2$ of all the agents, and (c) State $3$ of all the agents.}
    \label{fig:example1}
\end{figure}

The agents' dynamics exhibit marginal stability. Initially, each agent is assigned a random state within the range \([0, 10]\), without the application of any consensus law. To construct the matrices \( D_j \) and \( X_{Dj} \) for \( j = 1, \ldots, 5 \), each agent collects \( l+1 = 6 \) samples. A small Gaussian probing noise is introduced into the input signal to satisfy the persistency of excitation condition. After obtaining the 6 samples, the data-dependent optimization problem \eqref{eq:Hmudata} is solved, allowing for the derivation of the components of the optimal \( Q \)-function. Next, Algorithm \ref{alg:convex_concave} is utilized to compute the coupling gain. The overall computation time per agent was \( 4.45\,\text{s} \). The state trajectories of the agents, both during the model-free design process and after applying the distributed data-driven control at time-step 5, are presented in Figure \ref{fig:example1}. The results illustrate that dynamic consensus is successfully achieved.
\begin{figure}[ht]
    \centering
    \includegraphics[width=0.7\linewidth]{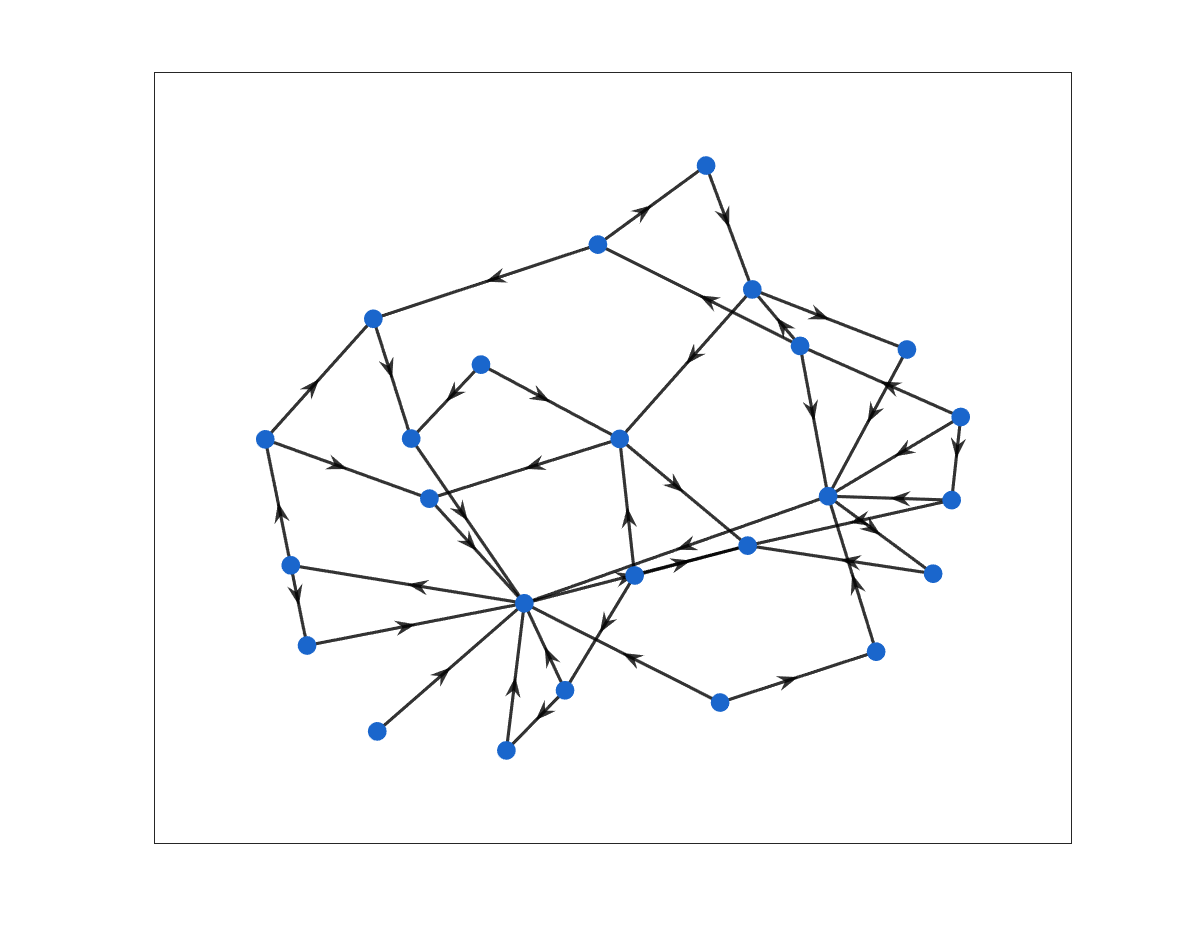}
    \caption{The communication graph in Example 2.}
    \label{fig:GExample2}
\end{figure}

\begin{figure}[ht]
    \centering
    \begin{subfigure}[t]{0.5\textwidth}
        \centering
        \includegraphics[width=0.75\textwidth]{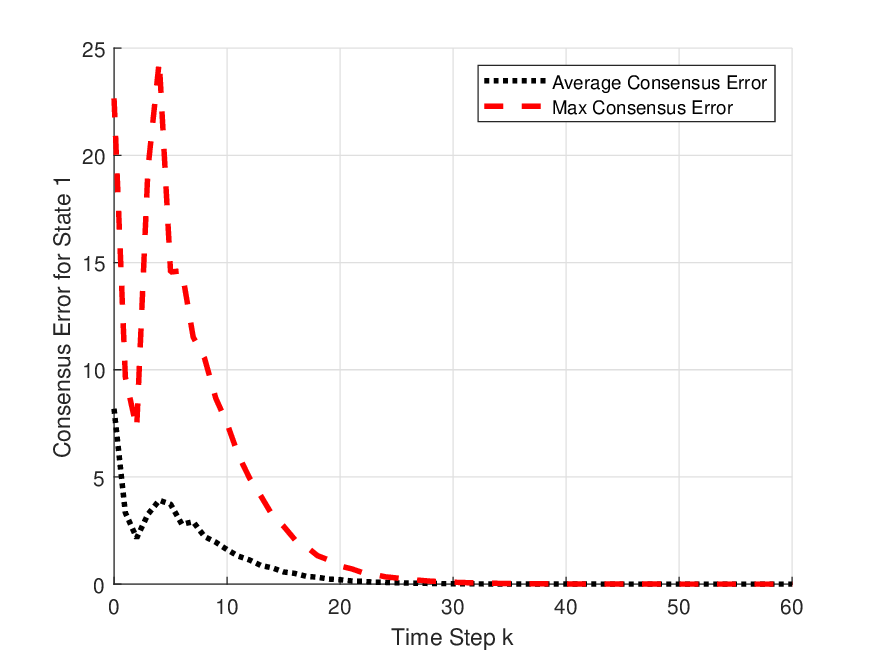}
        \caption{}
        \label{fig:subfig12}
    \end{subfigure}
    
    \begin{subfigure}[t]{0.5\textwidth}
        \centering
        \includegraphics[width=0.75\textwidth]{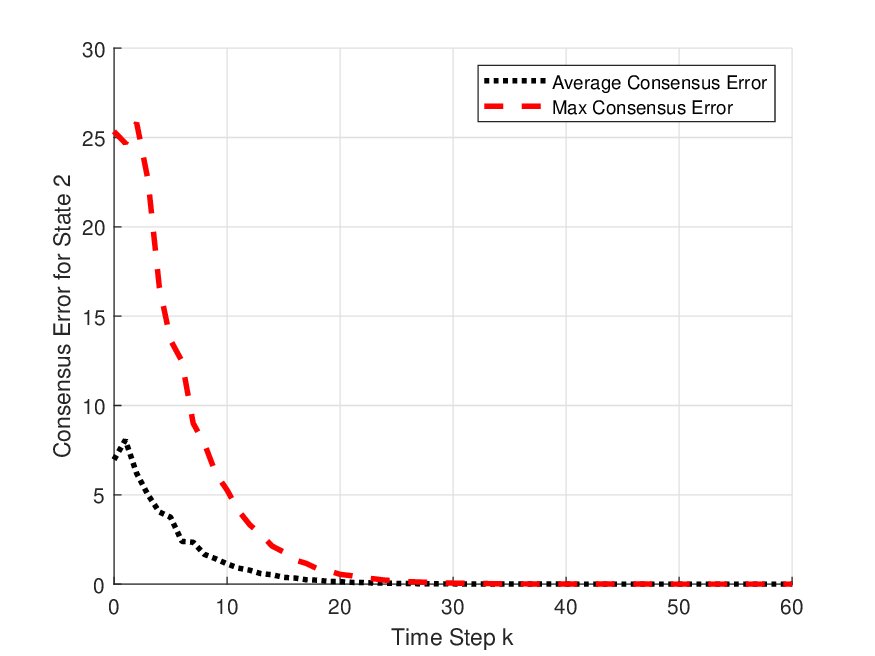}
       \caption{}
        \label{fig:subfig22}
    \end{subfigure}
    
    \begin{subfigure}[t]{0.5\textwidth}
        \centering
        \includegraphics[width=0.75\textwidth]{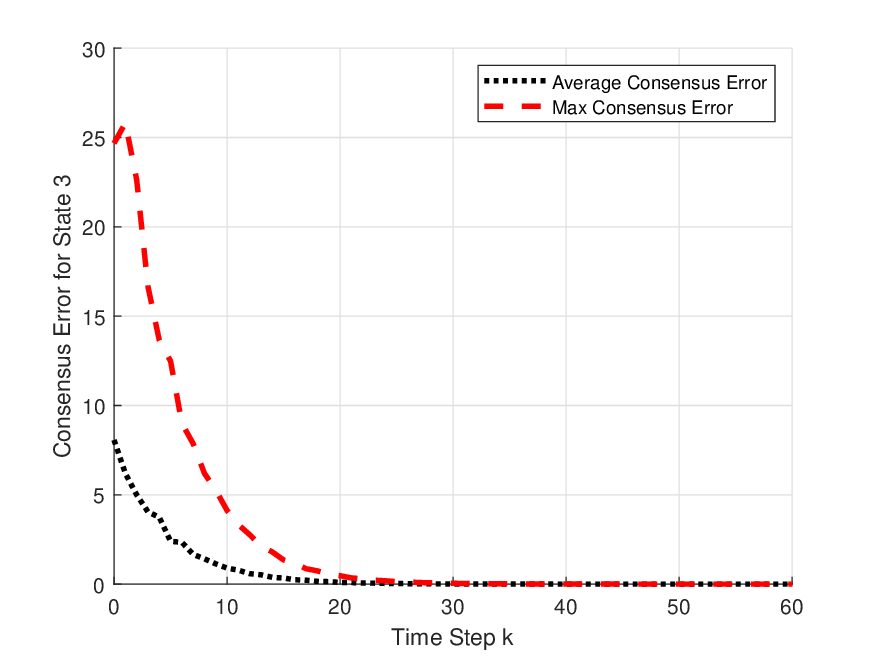}
       \caption{}
        \label{fig:subfig32}
    \end{subfigure}

    \caption{Dynamic consensus of 25 agents in Example 2 with  (a) State $1$ of all the agents, (b) State $2$ of all the agents, and (c) State $3$ of all the agents.}
    \label{fig:example2}
\end{figure}

\subsection*{Example 2: Large Scale Multi-agent System}
In this example a multi-agent system with \( N = 25 \) agents are considered. The dynamics of each agent is given by,
\[
A = \begin{bmatrix}
0.4 & 0.8 & 0 \\
0 & 0 & 1 \\
-0.2 & 0.2 & 1.1
\end{bmatrix}, \quad
B = \begin{bmatrix}
0.5 & 1 \\
0 & 0 \\
1 & 0
\end{bmatrix}.
\]
Let $Q= 10I$ and $\gamma=1$, and $\mu=1.1$. Consider the connectivity graph $\mathcal{G}$ be given as Figure \ref{fig:GExample2}. 

The dynamics of the agents are unstable. The system starts from a random initial condition for each agent without incorporation
of a consensus law, the samples are collected by each agent to build the matrices $D_j$ and $X_{Dj}$ for $j=1, \ldots 25$, through an offline experiment. The data-dependent optimization problem is solved and the Algorithm \ref{alg:convex_concave} is applied to find the coupling gain. The overall computation time per agent was \( 5.89\,\text{s} \). After deriving the control parameters, they are applied to the multi-agent system for dynamic consensus starting from an arbitrary initial condition within the range \([0, 10]\). The average consensus errors and the maximum consensus errors for the first, second, and third state of 25 agents are shown in Figures \ref{fig:example2} (a), (b), and (c),  by dot lines and dash lines, respectively.

\subsection*{Example 3: Feasibility Analysis}
In this example the feasibility and performance of the proposed method is compared with the method in \cite{feng2021consensusability,feng2021q}. To be able to compare the results, we set $\mu=1$ in the proposed method. Consider the dynamic consensus problem for the dynamics

\[
A = \begin{bmatrix}
    0 & 1 & 0 \\
    0 & 0 & 1 \\
    -0.2 & 0.2 & 1.1
\end{bmatrix}, \quad
B = \begin{bmatrix}
    0 \\
    0 \\
    1
\end{bmatrix},
\]

with the laplacian matrix
\[
\mathcal{L} = \begin{bmatrix}
    2 & 0 & 0 & 0 & -2 \\
    -5 & 6 & -1 & 0 & 0 \\
    -1 & 0 & 1 & 0 & 0 \\
    0 & 0 & -3 & 3 & 0 \\
    0 & 0 & -1 & 0 & 1
\end{bmatrix},
\]
with $Q=I$. The feasibility of the coupling gain for different values of $\gamma$ are reported in Table \ref{tab:gamma1}. {Theorem 6} demonstrates feasibility for smaller $\gamma$ values ($0 \leq \gamma \leq 0.1$) but becomes infeasible as $\gamma$ increases to 10 and beyond. This indicates its sensitivity to larger values of $\gamma$, making it suitable for scenarios with less emphasis on the minimization of the control signal energy. In contrast, {Algorithm 1} consistently achieves feasibility across all tested $\gamma$ values (${0 \leq \gamma \leq 1000}$), highlighting its robustness and adaptability to varying conditions. Lastly, the method from {\cite{feng2021consensusability}} is infeasible for smaller $\gamma$ values but begins to show feasibility as $\gamma$ reaches 100, demonstrating its preference for designs with high values of $\gamma$, i.e., minimal energy consensus.
In summary, Algorithm 1 provides the most reliable performance across all scenarios, while Theorem 6 and \cite{feng2021consensusability} have more specific applicability, with the former excelling at low $\gamma$ values and the latter at higher values.

\section{Conclusion}
This paper introduces a novel approach for achieving dynamic consensus in linear discrete-time homogeneous multi-agent systems with marginally stable or unstable agent dynamics. Our method addresses the coupling gain feasibility problem using convex formulation with LMI constraints, offering a systematic solution to a key challenge in multi-agent system design. Additionally, an iterative algorithm is proposed that guarantees convergence to a feasible coupling gain when the primary non-convex constraints are met. The framework also extends to model-free setups through efficient data-driven schemes, achieving performance equivalent to model-based approaches without additional conservatism beyond that inherent to the original formulation. Finally, a customized algorithm balances feasibility, convergence rate, robustness, and energy efficiency, enhancing adaptability for diverse applications. These contributions provide a robust foundation for advancing consensus control in multi-agent systems. Numerical simulations validate
the method’s effectiveness in providing feasible consensus solutions for large-scale multi-agent systems. Extending this framework to heterogeneous agents, with distinct dynamics, presents significant challenges and promising directions for future research.

\bibliographystyle{unsrt}
\bibliography{MyReferences}

\begin{thebibliography}{10}

\bibitem{qin2016recent}
Jiahu Qin, Qichao Ma, Yang Shi, and Long Wang.
\newblock Recent advances in consensus of multi-agent systems: A brief survey.
\newblock {\em IEEE Transactions on Industrial Electronics}, 64(6):4972--4983, 2016.

\bibitem{xia2015structural}
Weiguo Xia, Ming Cao, and Karl~Henrik Johansson.
\newblock Structural balance and opinion separation in trust--mistrust social networks.
\newblock {\em IEEE Transactions on Control of Network Systems}, 3(1):46--56, 2015.

\bibitem{liang2019prescribed}
Hongjing Liang, Yanhui Zhang, Tingwen Huang, and Hui Ma.
\newblock Prescribed performance cooperative control for multiagent systems with input quantization.
\newblock {\em IEEE Transactions on cybernetics}, 50(5):1810--1819, 2019.

\bibitem{chen2019control}
Fei Chen, Wei Ren, et~al.
\newblock On the control of multi-agent systems: A survey.
\newblock {\em Foundations and Trends{\textregistered} in Systems and Control}, 6(4):339--499, 2019.

\bibitem{zhang2011lyapunov}
Hongwei Zhang, Frank~L Lewis, and Zhihua Qu.
\newblock Lyapunov, adaptive, and optimal design techniques for cooperative systems on directed communication graphs.
\newblock {\em IEEE Transactions on Industrial Electronics}, 59(7):3026--3041, 2011.

\bibitem{li2009consensus}
Zhongkui Li, Zhisheng Duan, Guanrong Chen, and Lin Huang.
\newblock Consensus of multiagent systems and synchronization of complex networks: A unified viewpoint.
\newblock {\em IEEE Transactions on Circuits and Systems I: Regular Papers}, 57(1):213--224, 2009.

\bibitem{dutta2022strict}
Maitreyee Dutta, Elena Panteley, Antonio Loria, and Srikant Sukumar.
\newblock Strict lyapunov functions for dynamic consensus in linear systems interconnected over directed graphs.
\newblock {\em IEEE Control Systems Letters}, 6:2323--2328, 2022.

\bibitem{zhang2020decoupling}
Jilie Zhang, Tao Feng, Huaguang Zhang, and Xiaomin Wang.
\newblock The decoupling cooperative control with dominant poles assignment.
\newblock {\em IEEE Transactions on Systems, Man, and Cybernetics: Systems}, 52(2):1205--1213, 2020.

\bibitem{feng2021consensusability}
Tao Feng, Jilie Zhang, Yin Tong, and Huaguang Zhang.
\newblock Consensusability and global optimality of discrete-time linear multiagent systems.
\newblock {\em IEEE Transactions on Cybernetics}, 52(8):8227--8238, 2021.

\bibitem{de2019formulas}
Claudio De~Persis and Pietro Tesi.
\newblock Formulas for data-driven control: Stabilization, optimality, and robustness.
\newblock {\em IEEE Transactions on Automatic Control}, 65(3):909--924, 2019.

\bibitem{berberich2020trajectory}
Julian Berberich and Frank Allg{\"o}wer.
\newblock A trajectory-based framework for data-driven system analysis and control.
\newblock In {\em 2020 European Control Conference (ECC)}, pages 1365--1370. IEEE, 2020.

\bibitem{trentelman2021informativity}
Harry~L Trentelman, Henk~J Van~Waarde, and M~Kanat Camlibel.
\newblock An informativity approach to the data-driven algebraic regulator problem.
\newblock {\em IEEE Transactions on Automatic Control}, 67(11):6227--6233, 2021.

\bibitem{van2020noisy}
Henk~J van Waarde, M~Kanat Camlibel, and Mehran Mesbahi.
\newblock From noisy data to feedback controllers: Nonconservative design via a matrix {S}-lemma.
\newblock {\em IEEE Transactions on Automatic Control}, 67(1):162--175, 2020.

\bibitem{farjadnasab2022model}
Milad Farjadnasab and Maryam Babazadeh.
\newblock Model-free {LQR} design by {Q}-function learning.
\newblock {\em Automatica}, 137:110060, 2022.

\bibitem{Sahel}
Sahel {Vahedi Noori} and Maryam Babazadeh.
\newblock A data-ensemble-based approach for sample-efficient {LQ} control of linear time-varying systems.
\newblock {\em Journal of the Franklin Institute}, 362(16), 2025.

\bibitem{yang2020cooperative}
Ruohan Yang, Lu~Liu, and Gang Feng.
\newblock Cooperative output tracking of unknown heterogeneous linear systems by distributed event-triggered adaptive control.
\newblock {\em IEEE Transactions on Cybernetics}, 52(1):3--15, 2022.

\bibitem{baldi2020distributed}
Simone Baldi, Ilario~A Azzollini, and Petros~A Ioannou.
\newblock A distributed indirect adaptive approach to cooperative tracking in networks of uncertain single-input single-output systems.
\newblock {\em IEEE Transactions on Automatic Control}, 66(10):4844--4851, 2020.

\bibitem{cao2023distributed}
Wenji Cao, Lu~Liu, and Gang Feng.
\newblock Distributed adaptive output consensus of unknown heterogeneous non-minimum phase multi-agent systems.
\newblock {\em IEEE/CAA Journal of Automatica Sinica}, 10(4):997--1008, 2023.

\bibitem{jiang2024fully}
Yi~Jiang, Lu~Liu, and Gang Feng.
\newblock Fully distributed adaptive control for output consensus of uncertain discrete-time linear multi-agent systems.
\newblock {\em Automatica}, 162:111531, 2024.

\bibitem{gusrialdi2017distributed}
Azwirman Gusrialdi and Zhihua Qu.
\newblock Distributed estimation of all the eigenvalues and eigenvectors of matrices associated with strongly connected digraphs.
\newblock {\em IEEE control systems letters}, 1(2):328--333, 2017.

\bibitem{zhang2023data}
Xiufeng Zhang, Gang Wang, and Jian Sun.
\newblock Data-driven control of consensus tracking for discrete-time multi-agent systems.
\newblock {\em Journal of the Franklin Institute}, 360(7):4661--4674, 2023.

\bibitem{feng2021q}
Tao Feng, Jilie Zhang, Yin Tong, and Huaguang Zhang.
\newblock {Q}-learning algorithm in solving consensusability problem of discrete-time multi-agent systems.
\newblock {\em Automatica}, 128:109576, 2021.

\bibitem{west.2001}
Douglas~Brent West et~al.
\newblock {\em Introduction to graph theory}, volume~2.
\newblock Prentice hall Upper Saddle River, 2001.

\bibitem{ong2021consensus}
Chong-Jin Ong and Bonan Hou.
\newblock Consensus of heterogeneous multi-agent system with input constraints.
\newblock {\em Automatica}, 134:109895, 2021.

\bibitem{sun2024adaptive}
Jian Sun, Hongbo Lei, Jianxin Zhang, Lei Liu, and Qihe Shan.
\newblock Adaptive consensus control of multiagent systems with an unstable high-dimensional leader and switching topologies.
\newblock {\em IEEE Transactions on Industrial Informatics}, 20(9):10946--10953, 2024.

\bibitem{lewis2009}
Frank~L. Lewis and Draguna Vrabie.
\newblock Reinforcement learning and adaptive dynamic programming for feedback control.
\newblock {\em IEEE Circuits and Systems Magazine}, 9:32--50, 2009.

\bibitem{bertsekas2012dynamic}
Dimitri Bertsekas.
\newblock {\em Dynamic programming and optimal control: Volume I}, volume~4.
\newblock Athena scientific, 2012.

\bibitem{hengster2013synchronization}
Kristian Hengster-Movric, Keyou You, Frank~L Lewis, and Lihua Xie.
\newblock Synchronization of discrete-time multi-agent systems on graphs using {R}iccati design.
\newblock {\em Automatica}, 49(2):414--423, 2013.

\bibitem{boyd2004convex}
Stephen~P Boyd and Lieven Vandenberghe.
\newblock {\em Convex optimization}.
\newblock Cambridge university press, 2004.

\bibitem{lipp2016variations}
Thomas Lipp and Stephen Boyd.
\newblock Variations and extension of the convex--concave procedure.
\newblock {\em Optimization and Engineering}, 17:263--287, 2016.

\bibitem{willems2005note}
Jan~C Willems, Paolo Rapisarda, Ivan Markovsky, and Bart~LM De~Moor.
\newblock A note on persistency of excitation.
\newblock {\em Systems \& Control Letters}, 54(4):325--329, 2005.

\bibitem{bitmead1985monotonicity}
Robert~R Bitmead, Michel~R Gevers, Ian~R Petersen, and R~John Kaye.
\newblock Monotonicity and stabilizability-properties of solutions of the riccati difference equation: Propositions, lemmas, theorems, fallacious conjectures and counterexamples.
\newblock {\em Systems \& Control Letters}, 5(5):309--315, 1985.

\end{thebibliography}

\end{document}